\documentclass[reprint, aps, prb]{revtex4-1}

\usepackage{amsmath}	% for math expressions
\usepackage{graphicx}	% Include figure files
\usepackage{dcolumn}	% Align table columns on decimal point
\usepackage{hyperref}	% add hypertext capabilities
\usepackage{multirow}	% for cells spanning multirow in tables

\begin{document}

\title{Uncertainty quantification of thermal conductivities from equilibrium molecular dynamics simulations} 
\author{Zuyuan Wang,$^{a, b}$ Salar Safarkhani,$^{a}$ Guang Lin,$^{a, c}$ and Xiulin Ruan$^{a, b}$}
%\author{Xiulin Ruan}
\email{ruan@purdue.edu}
\affiliation{$^a$School of Mechanical Engineering, Purdue University, West Lafayette, Indiana 47907, USA\\ $^b$The Birck Nanotechnology Center, Purdue University, West Lafayette, Indiana 47907, USA\\ $^c$Department of Mathematics, Purdue University, West Lafayette, Indiana 47907, USA}
\date{\today}	% It is always \today, today, but any date may be explicitly specified

%%%
%%% ABSTRACT
%%%
\begin{abstract}
Equilibrium molecular dynamics (EMD) simulations along with the Green-Kubo formula have been widely used to calculate lattice thermal conductivities. Previous studies, however, focused primarily on the calculated thermal conductivities, with the uncertainty of the thermal conductivities remaining poorly understood. In this paper, we study the quantification of the uncertainty by using solid argon, silicon, and germanium as model material systems, and examine the origin of the observed uncertainty. We find that the uncertainty increases with the upper limit of the correlation time, $t_{\text{corre, UL}}$, and decreases with the total simulation time, $t_{\text{total}}$, whereas the velocity initialization seed, simulation domain size, temperature, and type of material have minimal effects. The relative uncertainties of the thermal conductivities, $\sigma_{k_x}/k_{x, \text{ave}}$, for solid argon, silicon, and germanium under different simulation conditions all follow a similar trend, which can be fit with a ``universal'' square-root relation, as $\sigma_{k_x}/k_{x, \text{ave}} = 2(t_{\text{total}}/t_{\text{corre, UL}})^{-0.5}$. We have also conducted statistical analysis of the EMD-predicted thermal conductivities and derived a formula that correlates the relative error bound ($Q$), confidence level ($P$), $t_{\text{corre, UL}}$, $t_{\text{total}}$, and number of independent simulations ($N$). We recommend choosing $t_{\text{corre, UL}}$ to be 5--10 times the effective phonon relaxation time, $\tau_{\text{eff}}$, and choosing $t_{\text{total}}$ and $N$ based on the desired relative error bound and confidence level. This study provides new insights into understanding the uncertainty of EMD-predicted thermal conductivities. It also provides a guideline for running EMD simulations to achieve a desired relative error bound with a desired confidence level and for reporting EMD-predicted thermal conductivities.  
\end{abstract}

\maketitle

%%%
%%% INTRODUCTION
%%%
\section{\label{sec:introduction}Introduction}
Equilibrium molecular dynamics (EMD) simulation along with the Green-Kubo formula is an effective way to calculate lattice thermal conductivities.~\cite{Green1954, Kubo1957, McQuarrie2000, Schelling2002, McGaughey2004, Wang2015, Wang2016a, Wang2016b} In this method, the thermal conductivity is related to the integration of the heat current autocorrelation function (HCACF), as~\cite{McQuarrie2000}  
%
% 
% eq:green-kubo
\begin{equation}
k_{\alpha\beta}=\frac{V}{k_{\text{B}}T^2}\int_0^{\infty} \! \langle J_{\alpha}(0)J_{\beta}(t_{\text{corre}})\rangle \;
\text{d}t_{\text{corre}}, 
\label{eq:green-kubo}
\end{equation}
where $k_{\alpha\beta}$ is the $\alpha\beta^{\text{th}}$ component of the thermal conductivity tensor, $V$ is the volume of the material system, $k_{\text{B}}$ is the Boltzmann constant, $T$ is temperature, $t_{\text{corre}}$ is the heat current autocorrelation time, and $J_{\alpha}$ is the $\alpha^{\text{th}}$ component of the full heat current vector $\mathbf{J}$, which is typically computed as~\cite{Plimpton1995}
%
%
% eq:heat-current
\begin{equation}
\mathbf{J} = \frac{1}{V} \left( \sum_i \mathbf{v}_i \epsilon_i 
+ \sum_{i} \mathbf{S}_i \cdot \mathbf{v}_i \right),
\label{eq:heat-current}
\end{equation}
Here, $\mathbf{v}_i$, $\epsilon_i$, and $\mathbf{S}_i$ are the velocity, energy, and stress of atom $i$. In \texttt{LAMMPS},~\cite{Plimpton1995} a widely used, open-source molecular dynamics simulation package, the default heat current formula is based on Eq.~(\ref{eq:heat-current}) with the interatomic forces calculated from the per-atom stresses. Recently Fan \textit{et al.} reported new heat current formulas for many-body potentials, but the different heat current formulas are shown to affect mainly low-dimensional materials.~\cite{Fan2015} In theory, the $V$, integration upper limit, and heat current autocorrelation time in Eq.~(\ref{eq:green-kubo}) should all approach infinity to calculate the lattice thermal conductivity of bulk materials. In real practice, however, the $V$ is chosen to be of a finite size based on some domain size effect studies, the integration is carried out up to a finite upper limit, which we define as the upper limit of the correlation time, $t_{\text{corre, UL}}$, and the heat current autocorrelation is calculated up to a finite duration, which we define as the total simulation time, $t_{\text{total}}$. As a result, Equation~(\ref{eq:green-kubo}) becomes 
%
% 
% eq:green-kubo-2
\begin{equation}
k_{\alpha\beta}=\frac{V}{k_{\text{B}}T^2}\int_0^{t_{\text{corre, UL}}} \! \langle J_{\alpha}(0)J_{\beta}(t_{\text{corre}})\rangle |_{t_{\text{total}}} \; \text{d}t_{\text{corre}}.  
\label{eq:green-kubo-2}
\end{equation}
Although this method has been widely used to calculate the lattice thermal conductivity of many material systems, previous studies focused primarily on the thermal conductivity values or the average values from multiple independent simulations, with the uncertainty of the predicted thermal conductivities remaining poorly understood, as seen from the very limited investigations on it so far.~\cite{Angelikopoulos2012, Marepalli2014, Race2015} On the other hand, it is a common practice to report both the values and uncertainties of the predicted thermal conductivities from EMD simulations. A typical way of doing it is to run each simulation for multiple times (usually 3--12) and then calculate the average value as the thermal conductivity and the standard deviation as the uncertainty (plotted as error bars). This practice, however, often lacks consistency (or a well-defined guideline) because the values and uncertainties could vary greatly depending on how the simulations are conducted. In addition, it was pointed out that the uncertainty of the thermal conductivity from EMD simulations is about 20\%,~\cite{Schelling2002} for which no explanation was provided. Furthermore, when thermal conductivities from EMD simulations are compared with those from other sources (\textit{e.g.}, experiments or other simulation methods), it is often concluded that the agreement is good if the error bars overlap, but little is known about the information carried by the error bars. 

In this study, we conduct a systematic study on quantifying the uncertainty of thermal conductivities from EMD simulations. We consider solid argon, silicon, and germanium as model material systems, and study the effects of the velocity initialization seed, simulation domain size, upper limit of the correlation time ($t_{\text{corre, UL}}$), total simulation time ($t_{\text{total}}$), temperature, and type of material. The results show that the uncertainty increases with $t_{\text{corre, UL}}$ and decreases with $t_{\text{total}}$, but the velocity initialization seed, simulation domain size, temperature, and type of material have minimal effects on the relative uncertainty. By analyzing the results of different materials under different simulation conditions, we have obtained a ``universal'' square-root relation for quantifying the relative uncertainty, $\sigma_k/k_{\text{ave}}$, as a function of $t_{\text{total}}/t_{\text{corre, UL}}$. We have also obtained a formula that correlates the relative error bound ($Q$), confidence level ($P$), $t_{\text{corre, UL}}$, $t_{\text{total}}$, and number of independent simulations ($N$). This paper is organized as follows. Section~\ref{sec:methodology} details the method used in this study, particularly the EMD simulations. Section~\ref{sec:results-discussion} presents some results and discussion on the uncertainty of the EMD-predicted thermal conductivities of solid argon, silicon, and germanium. It also reports some results on quantifying the general uncertainty of EMD-predicted thermal conductivities, choosing appropriate $t_{\text{corre, UL}}$ and $t_{\text{total}}$ for EMD simulations to achieve a desired relative error bound with a desired confidence level, and reporting EMD-predicted thermal conductivities. Section~\ref{sec:conclusions} summarizes the main findings from this study.

%%%
%%% METHODOLOGY -- EMD simulations
%%%
\section{\label{sec:methodology}Methodology}
All the molecular dynamics simulations were conducted with the \texttt{LAMMPS} package.~\cite{Plimpton1995} The material systems of solid argon, silicon, and germanium all have a face-centered cubic (FCC) structure with nominal lattice constants (before the structures are relaxed) of 5.26, 5.43, and 5.66~\AA, respectively. The interatomic interactions are characterized with the Lennard-Jones potential~\cite{Allen1987} for solid argon and the Tersoff potential~\cite{Tersoff1988} for silicon and germanium. We considered a domain size of $6\times6\times6$ unit cells (u.c.) for solid argon (except for the domain size effect studies) and $4\times4\times4$~u.c. for silicon and germanium. Periodic boundary conditions were applied in $x$, $y$, and $z$ directions. The time steps were chosen as 4, 1, and 2~fs for solid argon, silicon, and germanium, respectively. Nos\'{e}-Hoover barostat and thermostat~\cite{Nose1984, Hoover1985} were used to control the pressure and temperature of the material systems. In all simulations, the material systems were first equilibrated in an $NPT$ (constant number of atoms, pressure, and temperature) ensemble before they were switched to an $NVE$ (constant number of atoms, volume, and energy) ensemble for data production. The $t_{\text{corre, UL}}$ and $t_{\text{total}}$ values were chosen such that the predicted average thermal conductivities converged. We varied the $t_{\text{corre, UL}}$ and $t_{\text{total}}$ over a wide range to investigate their effects on the uncertainty of the predicted thermal conductivities. Each simulation was run for 100 times, which had independent initial velocity distributions. It is an inherent assumption in this study that 100 independent simulations provide a representative sample for the relevant statistical analysis. The thermal conductivities were calculated according to~\hyperref[eq:green-kubo-2]{Eq.~(\ref{eq:green-kubo-2})}. Since each individual EMD simulation can provide three thermal conductivity values (for the $x$, $y$, and $z$ directions), there are a total of 300 thermal conductivity values for each simulation condition. We report the average and standard deviation of the 300 values as the predicted thermal conductivity and its uncertainty, respectively. Because the three materials considered in this study are all isotropic in the $x$, $y$, and $z$ directions, the 100 $k_x$, 100 $k_y$, and 100 $k_z$ values for each simulation condition could be equivalently treated as 300 $k_x$ values. As a results, the analysis in this study essentially corresponds to the thermal conductivity along a single direction. Alternatively, the thermal conductivities can be first averaged over the $x$, $y$, and $z$ directions and then the average and standard deviation of the 100 values from the 100 simulations calculated as the predicted thermal conductivity and its uncertainty, respectively. The average from these two methods will be the same, but the standard deviation from the second method will be statistically ${1}/{\sqrt{3}}$ times that from the first method. Considering the second method is restricted to isotropic materials, we adopted the first method to make our analysis more general.

%%%
%%% RESULTS AND DISCUSSION
%%%
\section{\label{sec:results-discussion}Results and discussion}
In this section, we present the results for the three materials --- solid argon, silicon, and germanium. For solid argon, we show the effects of the velocity initialization seed, simulation domain size, $t_{\text{total}}$, $t_{\text{corre, UL}}$, and temperature on the EMD-predicted thermal conductivity and its uncertainty. For silicon and germanium, we focus on the effects of the $t_{\text{total}}$ and $t_{\text{corre, UL}}$. Based on the solid argon, silicon, and germanium results, we provide some consideration on quantifying of the general uncertainty of EMD-predicted thermal conductivities. We also show how to appropriately choose $t_{\text{corre, UL}}$, $t_{\text{total}}$, and $N$ for EMD simulations so that the predicted average thermal conductivity achieves a desired relative error bound with a desired confidence level and how to report EMD-predicted thermal conductivities.

%%%
%%% RESULTS --- Solid Argon
%%% 
\subsection{\label{sec:results-solid-argon}Solid Argon}
%
% 
% fig:Fig1_Ar_40K_HCACF_k_size_effect
\begin{figure}[tbp]
\centering
\includegraphics[width=0.45\textwidth]{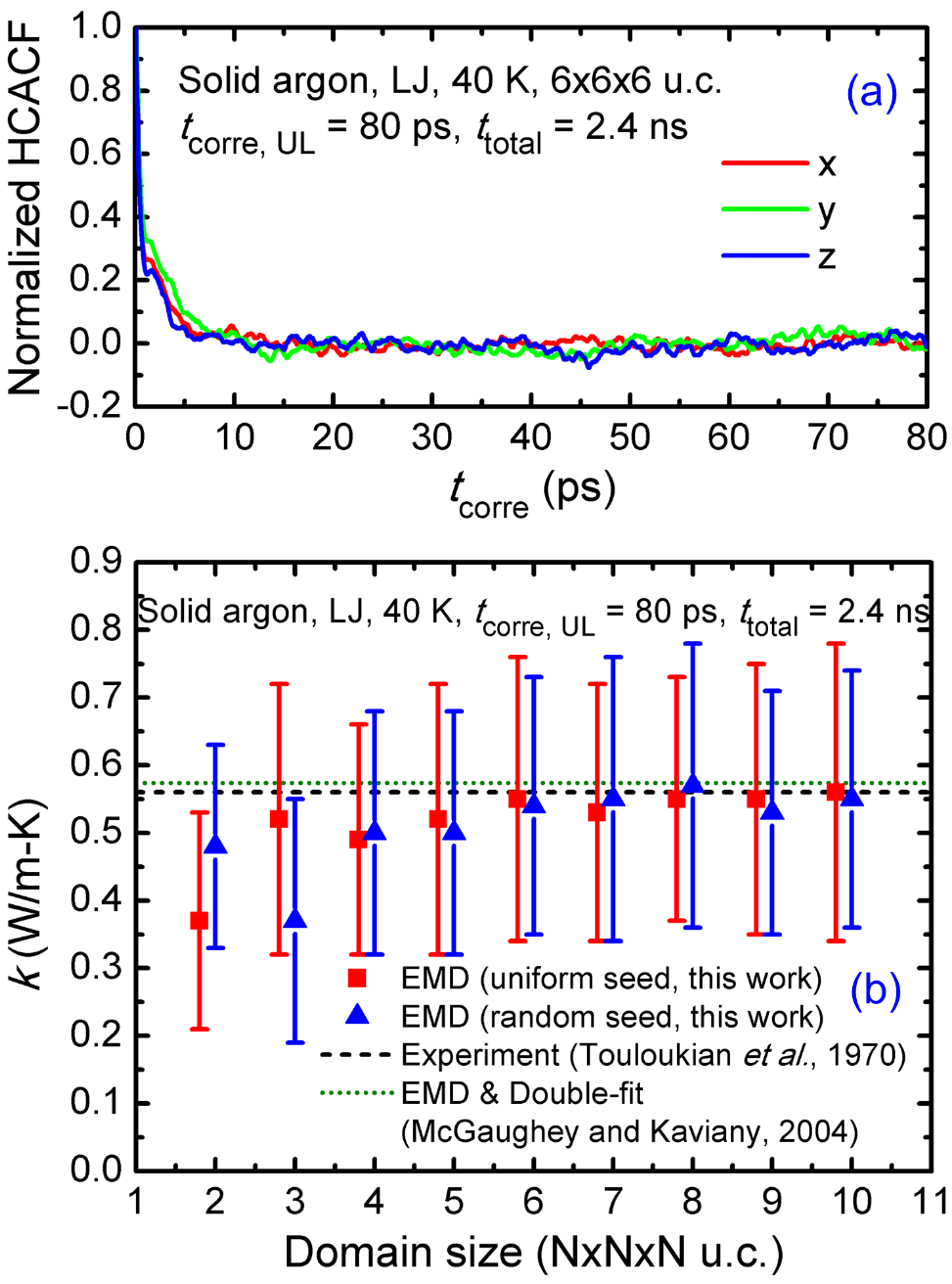} 
\caption{(Color online) (a) Typical heat current autocorrelation function (HCACF) profiles of solid argon, normalized by the initial HCACF values. (b) Variation of the EMD-predicted thermal conductivity of solid argon at 40~K with the simulation domain size. Each data point shows the results of 100 independent simulations (or 300 values). The data points for the uniform seeds are shifted slightly to the left for better readability. The black (dashed) and green (dotted) lines show the experimental result (0.560~W/m-K) by Touloukian \textit{et al.}~\cite{Touloukian1970} and simulation result (0.574~W/m-K) by McGaughey and Kaviany,~\cite{McGaughey2004} respectively. }
\label{fig:Fig1_Ar_40K_HCACF_k_size_effect}
\end{figure}
%
%
%
%
% fig:Fig2_Ar_k_distribution_velocity_seed_effect
\begin{figure*}[tbp]
\centering
\includegraphics[width=0.95\textwidth]{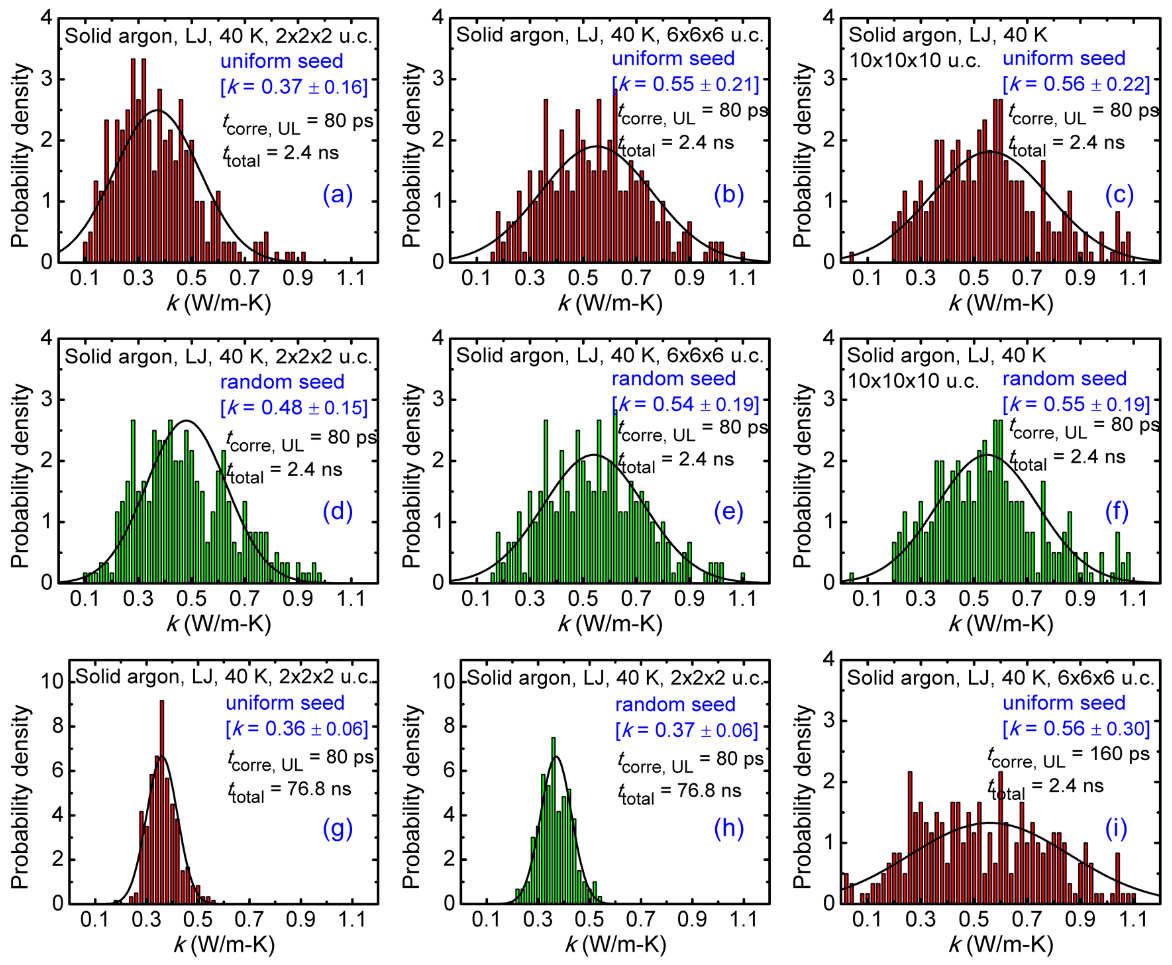} 
\caption{(Color online) Histogram distributions of thermal conductivities of solid argon from 100 independent simulations (or 300 values) with uniform or random velocity initialization seeds. The simulation conditions are indicated in the subfigures. The bell-shaped curves in the subfigures represent normal distribution curves with the same average thermal conductivity and uncertainty. }
\label{fig:Fig2_Ar_k_distribution_velocity_seed_effect}
\end{figure*}
%
%
%
%
% fig:Fig3_Ar_40K_6x6x6uc_ttotal_effect_on_HCACF_k_distribution
\begin{figure*}[tbp]
\centering
\includegraphics[width=0.9\textwidth]{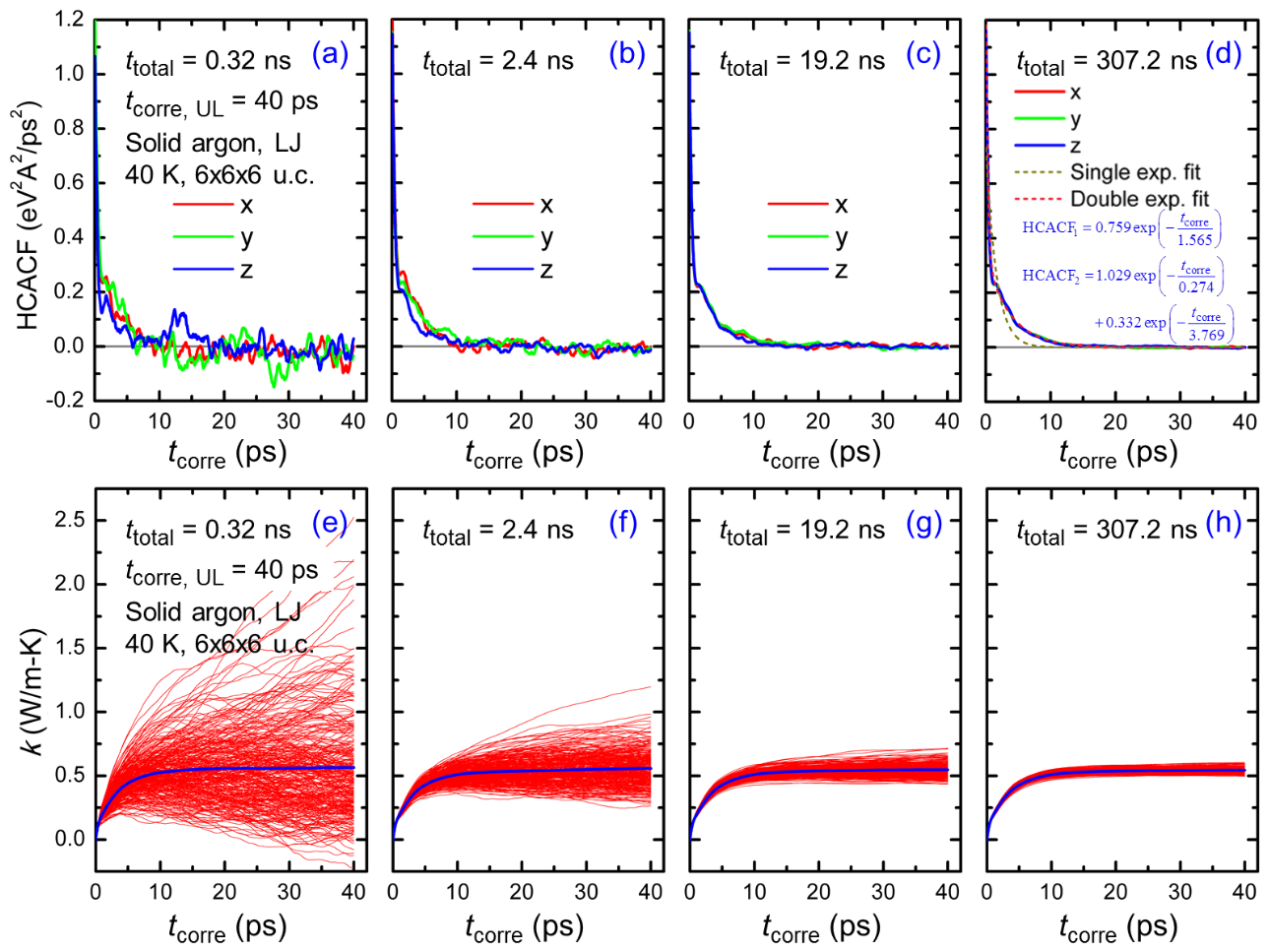} 
\caption{(Color online) (a)--(d) Typical HCACF profiles of solid argon with a domain size of $6\times6\times6$~u.c. at 40~K. The simulations correspond to $t_{\text{corre, UL}} = 40$~ps but different $t_{\text{total}}$ values. In (d) the results based on single- and double-exponential fitting are included to show how the effective phonon relaxation time, $\tau_{\text{eff}}$, could be obtained. (e)--(h) Distributions of thermal conductivity integration profiles of solid argon. The simulation conditions are indicated in the subfigures. The red (thin) curves represent the thermal conductivities from 100 independent simulations (including a total of 300 curves). The blue (thick) curves represent the corresponding average thermal conductivities. }
\label{fig:Fig3_Ar_40K_6x6x6uc_ttotal_effect_on_HCACF_k_distribution}
\end{figure*}
In EMD simulations, independent simulations are usually conducted to reduce the statistical error, which can be realized by assigning different velocity initialization seeds. In LAMMPS, the only requirement for a velocity initialization seed is that it be a positive integer.~\cite{Plimpton1995} To understand how the seeds affect the thermal conductivity predictions, we considered two schemes of assigning the seeds, namely, uniform and random seeds. The uniform seeds are described as $1000n$, where $n$ is the simulation ID (varying from 1 to 100), whereas the random seeds are random numbers (from 1000 to 100000) generated with the \texttt{rand} function of \texttt{MATLAB}. In Fig.~\ref{fig:Fig1_Ar_40K_HCACF_k_size_effect}(a), we show some typical HCACF profiles for solid argon. It is seen that the normalized HCACF starts from one, decreases gradually to zero, and then fluctuates around zero. Typically the correlation time, $t_{\text{corre, UL}}$, should be long enough so that the HCACF profiles cross zero for multiple times. The results in Fig.~\ref{fig:Fig1_Ar_40K_HCACF_k_size_effect}(a) indicates that $t_{\text{corre, UL}} = 80$~ps is sufficiently long for calculating the thermal conductivity of solid argon at 40~K. The discrepancy of the HCACF profiles in $x$, $y$, and $z$ directions could be attributed to the finite domain size, finite total simulation time, and statistical nature of molecular dynamics simulations. In Fig.~\ref{fig:Fig1_Ar_40K_HCACF_k_size_effect}(b), we show the domain size effect of the EMD-predicted thermal conductivity of solid argon, including the results with both the uniform and random seeds. Note that the data points for the uniform seeds are shifted slightly to the left for better readability. It is seen that the thermal conductivity increases with the increasing domain size and converges at a size around $6\times6\times6$~u.c. The results with uniform and random seeds differ appreciably for small domain sizes, but they agree reasonably well for domain sizes larger than $6\times6\times6$~u.c. The relative uncertainty of the thermal conductivities is around 25\%. We will show that this level of uncertainty comes primarily from the choices of $t_{\text{corre, UL}}$ and $t_{\text{total}}$. Upon convergence, the average thermal conductivities with both the uniform and random seeds agree well with the experimental value (0.560~W/m-K) by Touloukian \textit{et al.}~\cite{Touloukian1970} and the simulation result (0.574~W/m-K) by McGaughey and Kaviany.~\cite{McGaughey2004} 

In Fig.~\ref{fig:Fig2_Ar_k_distribution_velocity_seed_effect}, we show the detailed thermal conductivity distributions corresponding to the different simulation conditions. It is seen that the histogram distributions of the thermal conductivities under different simulation conditions all agree reasonably well with the corresponding normal distribution curves. Comparing the results in Fig.~\ref{fig:Fig2_Ar_k_distribution_velocity_seed_effect}(a) and (d), we notice that at a domain size of $2\times2\times2$~u.c., $t_{\text{corre, UL}} = 80$~ps, and $t_{\text{total}} = 2.4$~ns, significant discrepancies exist between the calculated average thermal conductivity with the uniform seeds and that with the random seeds. As the simulation domain size increases, the discrepancy decreases (see Fig.~\ref{fig:Fig2_Ar_k_distribution_velocity_seed_effect}(b) and (e) or Fig.~\ref{fig:Fig2_Ar_k_distribution_velocity_seed_effect}(c) and (f)), which could be attributed to the more ergodic sampling of the phase space at a larger domain size. The standard deviations of the thermal conductivity distributions are comparable for domain sizes of $6\times6\times6$ and $10\times10\times10$ u.c., suggesting a weak dependence of the standard deviation on the domain size upon convergence. We also considered the effects of $t_{\text{total}}$ and $t_{\text{corre, UL}}$. As shown in Fig.~\ref{fig:Fig2_Ar_k_distribution_velocity_seed_effect}(g) and (h), the average and standard deviation of the thermal conductivities with the uniform and random seeds agree well at $t_{\text{total}} = 76.8$~ns, even at a small domain size of $2\times2\times2$~u.c. This implies that the limited sampling of the phase space due to the limited domain size could be compensated by using a long total simulation time. Comparing the results in Fig.~\ref{fig:Fig2_Ar_k_distribution_velocity_seed_effect}(b) and (i), we observe a similar average thermal conductivity but a much flatter distribution (meaning a larger standard deviation) of the thermal conductivities, as $t_{\text{corre, UL}}$ increases from 80 to 160~ps. 

After noticing the significant effects of $t_{\text{total}}$ and $t_{\text{corre, UL}}$, we conducted a more detailed study on their effects on the predicted average thermal conductivity and the uncertainty. Figure~\ref{fig:Fig3_Ar_40K_6x6x6uc_ttotal_effect_on_HCACF_k_distribution} shows the HCACF and thermal conductivity integration profiles for solid argon at 40~K, which correspond to a domain size of $6\times6\times6$~u.c. and $t_{\text{corre, UL}} = 40$~ps but different $t_{\text{total}}$ values. The value for $t_{\text{corre, UL}}$ was chosen based on some preliminary simulation results (see Fig.~\ref{fig:Fig1_Ar_40K_HCACF_k_size_effect}(a)). We will provide some guidance on choosing an appropriate $t_{\text{corre, UL}}$ in Sec.~\ref{sec:choice-of-tcorreUL}. As $t_{\text{total}}$ increases, the HCACF profiles become smoother, and the difference between the HCACF profiles in different directions decreases. As a result, the distribution of the thermal conductivities becomes increasingly concentrated, as $t_{\text{total}}$ increases. We point out that the commonly seen fluctuations of the HCACF profiles around zero is non-intrinsic, because they could be reduced and eventually eliminated by increasing the total simulation time. For the thermal conductivity distribution, when $t_{\text{total}}$ is very small, abnormal thermal conductivities (such as negative or extremely large values) could appear; when $t_{\text{total}}$ is large, the thermal conductivities from the independent simulations form a narrow band, approaching a single value in the theoretical limit. Regarding the average thermal conductivity, all simulation conditions result in a similar value, which, to some extent, demonstrates the equivalence of time-averaging and ensemble-averaging.~\cite{Gordiz2015} 

\begin{figure}[tbp]
\centering
\includegraphics[width=0.45\textwidth]{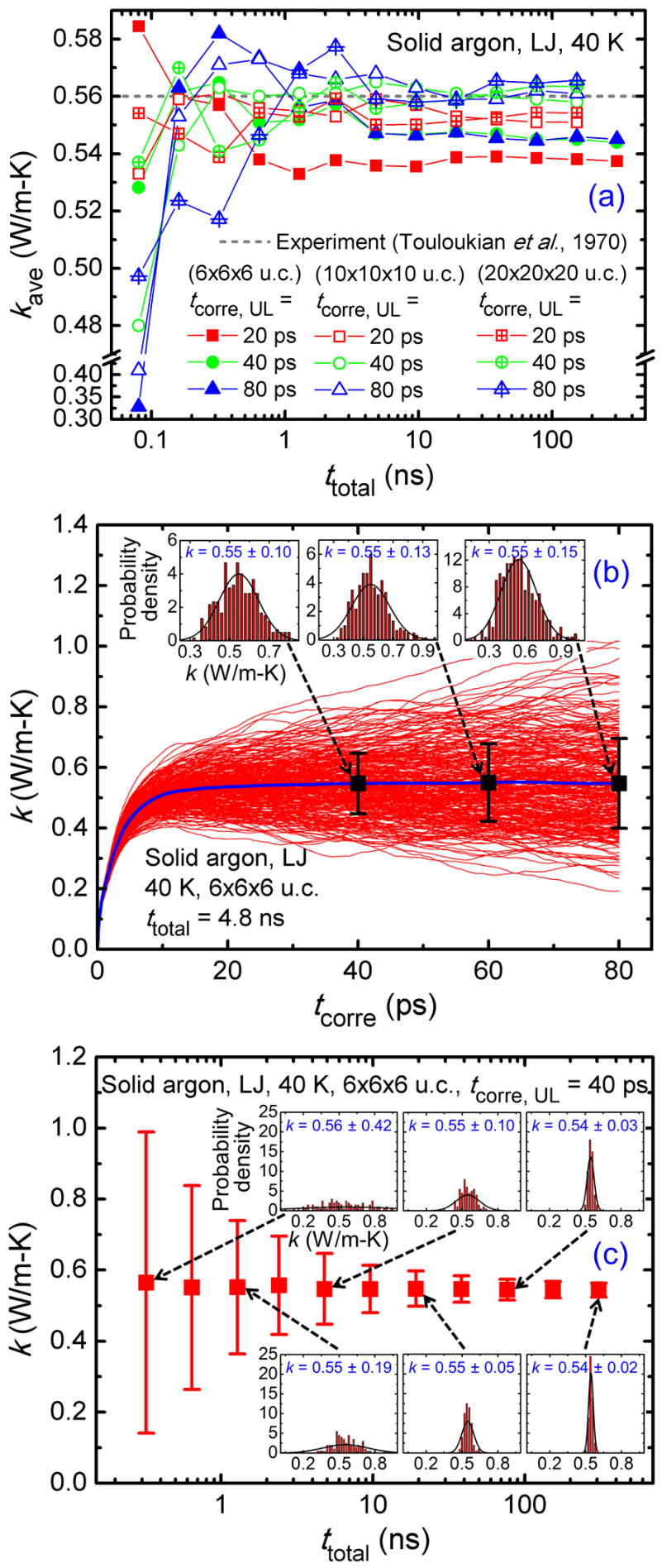} 
\caption{(Color online) (a) Variation of the EMD-predicted average thermal conductivity of solid argon at 40~K with $t_{\text{total}}$. The experimental value (0.560 W/m-K) by Touloukian \textit{et al.}~\cite{Touloukian1970} is shown as a comparison. (b) Variation of the thermal conductivity of solid argon at 40~K with $t_{\text{corre, UL}}$. (c) Variation of the thermal conductivity of solid argon at 40~K with $t_{\text{total}}$. The insets in (b) and (c) show the histogram distributions of the thermal conductivities corresponding to the simulation conditions. The bell-shaped curves represent normal distribution curves with the same average thermal conductivity and uncertainty. }
\label{fig:Fig4_Ar_40K_tcorre_total_effects_on_kave_and_dk}
\end{figure}

We consider further the effects of $t_{\text{total}}$ and $t_{\text{corre, UL}}$ on the EMD-predicted average thermal conductivity of solid argon and the uncertainty. Figure~\ref{fig:Fig4_Ar_40K_tcorre_total_effects_on_kave_and_dk}(a) shows the variation of the average thermal conductivity with $t_{\text{total}}$ at three $t_{\text{corre, UL}}$ values. When $t_{\text{total}}$ is short, the average thermal conductivity differs greatly from the experimental value. When $t_{\text{total}}$ is longer than a few nanoseconds, the average thermal conductivity becomes stable. The converged average thermal conductivity generally increases with $t_{\text{corre, UL}}$, with the value in good agreement with the experimental one for $t_{\text{total}}$ longer than 40~ps. This confirms the appropriateness of the choice of $t_{\text{corre, UL}}$ for the results shown in  Fig.~\ref{fig:Fig3_Ar_40K_6x6x6uc_ttotal_effect_on_HCACF_k_distribution}. Figure~\ref{fig:Fig4_Ar_40K_tcorre_total_effects_on_kave_and_dk}(b) shows the variation of the thermal conductivity distribution with $t_{\text{corre}}$. As $t_{\text{corre}}$ increases, the distribution expands, indicating a larger uncertainty of the predicted thermal conductivities. The insets show the histogram distributions of the thermal conductivities corresponding to $t_{\text{corre, UL}} = 40$, 60, and 80~ps, respectively, in comparison with the normal distribution curves.  Figure~\ref{fig:Fig4_Ar_40K_tcorre_total_effects_on_kave_and_dk}(c) shows the variation of the thermal conductivity distribution with $t_{\text{total}}$. As $t_{\text{total}}$ increases, the average remains nearly constant, but the error bars become smaller, indicating a smaller uncertainty of the predicted thermal conductivities. The insets show the histogram distributions of the thermal conductivities corresponding to different $t_{\text{total}}$ values, in comparison with the normal distribution curves. As $t_{\text{total}}$ increases, the histogram distribution becomes more and more peaked, confirming the decreasing uncertainty of the predicted thermal conductivities.  

\begin{figure}[tbp]
\centering
\includegraphics[width=0.45\textwidth]{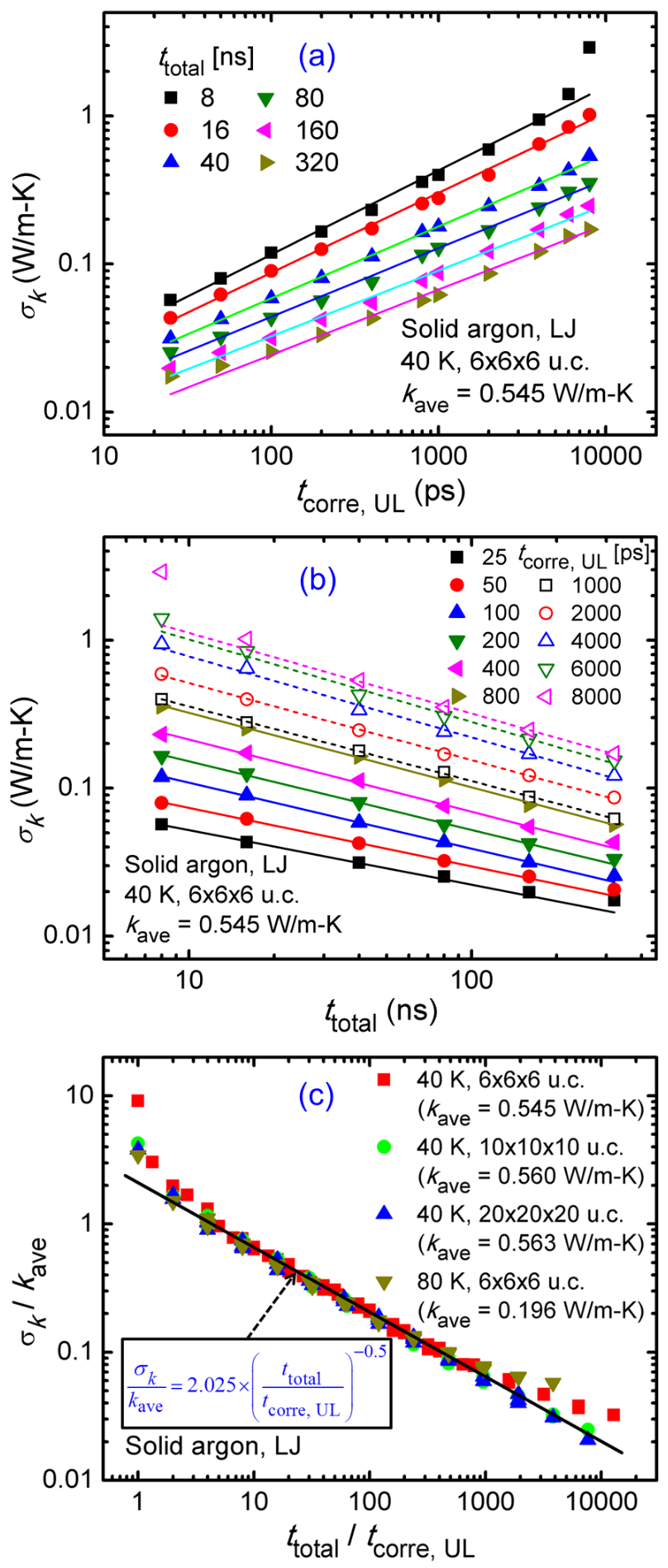} 
\caption{(Color online) (a) Variation of the uncertainty of the EMD-predicted thermal conductivity of solid argon with $t_{\text{corre, UL}}$. (b) Variation of the uncertainty of the thermal conductivity of solid argon with $t_{\text{total}}$. The results in (a) and (b) correspond to a solid argon material system with a domain size of $6\times6\times6$~u.c. at 40~K. (c) Variation of the relative uncertainty of the thermal conductivity of solid argon with $t_{\text{total}}/t_{\text{corre, UL}}$ corresponding to three domain sizes and two temperatures. The formula in (c) represents a fit of the data with a square-root relation. }
\label{fig:Fig5_Ar_ttotal_tcorre_effects_on_dk}
\end{figure}

Having seen the negligible effects of $t_{\text{total}}$ and $t_{\text{corre, UL}}$ on the EMD-predicted average thermal conductivity of solid argon and their qualitative effects on the uncertainty, we move on to quantitatively understand the effects of $t_{\text{total}}$ and $t_{\text{corre, UL}}$ on the uncertainty of the EMD-predicted thermal conductivity of solid argon. 
In Fig.~\ref{fig:Fig5_Ar_ttotal_tcorre_effects_on_dk}(a), we show the variation of the standard deviation of the thermal conductivities with $t_{\text{corre, UL}}$. For all $t_{\text{total}}$ values, $\sigma_k$ increases as $t_{\text{corre, UL}}$ increases. We fit the data with a power law relation as $\sigma_k \sim t_{\text{corre, UL}}^{\alpha}$, and the resulted $\alpha$ varied from 0.48 to 0.52. In Fig.~\ref{fig:Fig5_Ar_ttotal_tcorre_effects_on_dk}(b), we show the variation of the standard deviation of the thermal conductivities with $t_{\text{total}}$. For all $t_{\text{corre, UL}}$ values, $\sigma_k$ decreases as $t_{\text{total}}$ increases. We fit the data with a power law relation as $\sigma_k \sim t_{\text{corre, UL}}^{-\beta}$, and the resulted $\beta$ again varied from 0.48 to 0.52. 
%
%
% fig:Fig4_Ar_40K_tcorre_total_effects_on_kave_and_dk
%
%
%
%
% fig:Fig5_Ar_ttotal_tcorre_effects_on_dk
%
%  
%
%
% fig:Fig6_Si_500K_4x4x4uc_ttotal_effect_on_HCACF_k_distribution
\begin{figure*}[tbp]
\centering
\includegraphics[width=0.9\textwidth]{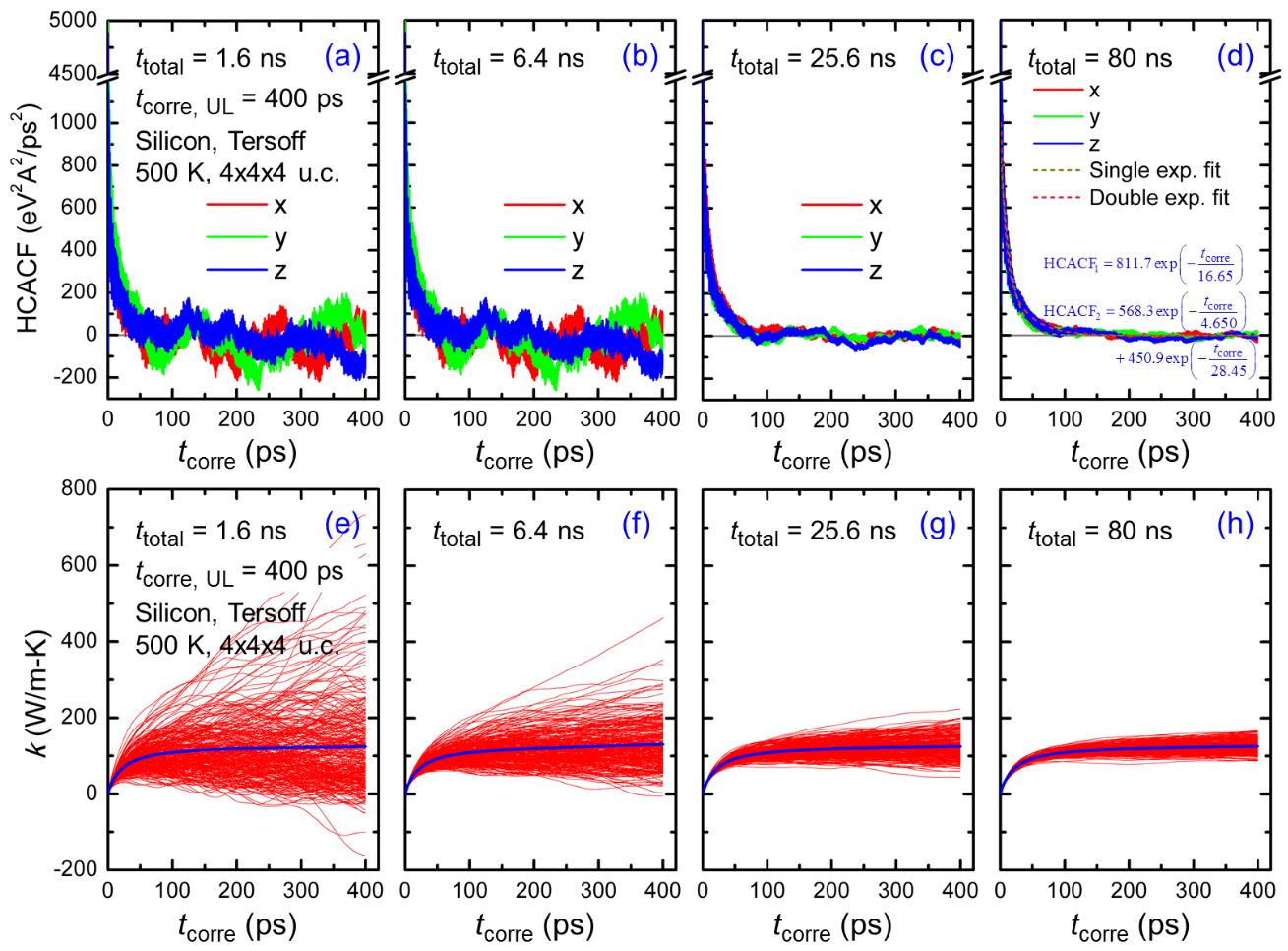} 
\caption{(Color online) (a)--(d) Typical HCACF profiles of silicon with a domain size of $4\times4\times4$~u.c. at 500~K. The simulations correspond to $t_{\text{corre, UL}} = 400$~ps but different $t_{\text{total}}$ values. In (d) the results based on single- and double-exponential fitting are included to show how the effective phonon relaxation time, $\tau_{\text{eff}}$, could be obtained. (e)--(h) Distributions of thermal conductivity integration profiles of silicon. The simulation conditions are indicated in the subfigures. The red (thin) curves represent the thermal conductivities from 100 independent simulations (including a total of 300 curves). The blue (thick) curves represent the corresponding average thermal conductivities.}
\label{fig:Fig6_Si_500K_4x4x4uc_ttotal_effect_on_HCACF_k_distribution}
\end{figure*}
These results suggests a square-root relation as $\sigma_k/k_{\text{ave}} \sim (t_{\text{total}}/t_{\text{corre, UL}})^{-0.5}$, where the average thermal conductivity, $k_{\text{ave}}$, is used to nondimensionalize $\sigma_k$. After plotting $\sigma_k/k_{\text{ave}}$ with $t_{\text{total}}/t_{\text{corre, UL}}$, we realize that the square-root relation indeed holds, especially for $t_{\text{total}}/t_{\text{corre, UL}}$ in the range from 5 to 2000, as seen in Fig.~\ref{fig:Fig5_Ar_ttotal_tcorre_effects_on_dk}(d). To confirm that the relation holds true for other domain sizes and temperatures, we conducted some additional EMD simulations with domain sizes of $10\times10\times10$ and $20\times20\times20$~u.c. and at another temperature 80~K. Surprisingly, all data follows a similar trend, as seen in Fig.~\ref{fig:Fig5_Ar_ttotal_tcorre_effects_on_dk}(d). We fit all the data with the square-root relation, and the result shows $\sigma_k/k_{\text{ave}} = 2.025(t_{\text{total}}/t_{\text{corre, UL}})^{-0.5}$. As a test of this relation, we consider the EMD results in Fig.~\ref{fig:Fig1_Ar_40K_HCACF_k_size_effect}(b). It turns out that the predicted relative uncertainty ($\sigma_k/k_{\text{ave}} = 37\%$) by the square-root relation is in excellent agreement with the actual relative uncertainty (around 36\%). In the following sections, we will show that this kind of square-root relation is not limited to solid argon.

%%%
%%% RESULTS --- Silicon
%%%
\subsection{\label{sec:results-silicon}Silicon}
Considering the significant effects of $t_{\text{total}}$ and $t_{\text{corre, UL}}$ on the uncertainty of EMD-predicted thermal conductivities of solid argon, we focus on the effects of these two parameters in the study on the silicon material system. In Fig.~\ref{fig:Fig6_Si_500K_4x4x4uc_ttotal_effect_on_HCACF_k_distribution}, we show the HCACF and thermal conductivity integration profiles for silicon at 500~K, which correspond to a domain size of $4\times4\times4$~u.c. and $t_{\text{corre, UL}} = 400$~ps but different $t_{\text{total}}$ values. Similar to the solid argon results, the value of $t_{\text{corre, UL}}$ was chosen based on some preliminary simulation results. As $t_{\text{total}}$ increases, the HCACF profiles become smoother, and the difference between the HCACF profiles in different directions decreases. As a result, the distribution of the thermal conductivities becomes more concentrated, as $t_{\text{total}}$ increases. Compared with the results for solid argon (see Fig.~\ref{fig:Fig3_Ar_40K_6x6x6uc_ttotal_effect_on_HCACF_k_distribution}), the fluctuations of the HCACF profiles for silicon is much larger, which could be attributed to the optical phonons as there are two basis atoms in a primitive unit cell of silicon. Still, the fluctuations are considered non-intrinsic, because they could be reduced and eventually eliminated by increasing the total simulation time. For the thermal conductivity distribution, when $t_{\text{total}}$ is very small, abnormal thermal conductivities (such as negative or extremely large values) could appear; when $t_{\text{total}}$ is large, the thermal conductivities from the independent simulations form a narrow band, approaching a single value in the theoretical limit. Regarding the average thermal conductivity, all simulation conditions result in a similar value, which again demonstrates the equivalence of time-averaging and ensemble-averaging.~\cite{Gordiz2015} The results for silicon are similar to those for solid argon, except for the different HCACF and thermal conductivity values, which arise primarily from the different atomic masses and interatomic potentials of the two materials. 
%
%
% fig:Fig7_Si_tcorre_ttotal_effects_on_kave_and_dk
\begin{figure}[tbp]
\centering
\includegraphics[width=0.45\textwidth]{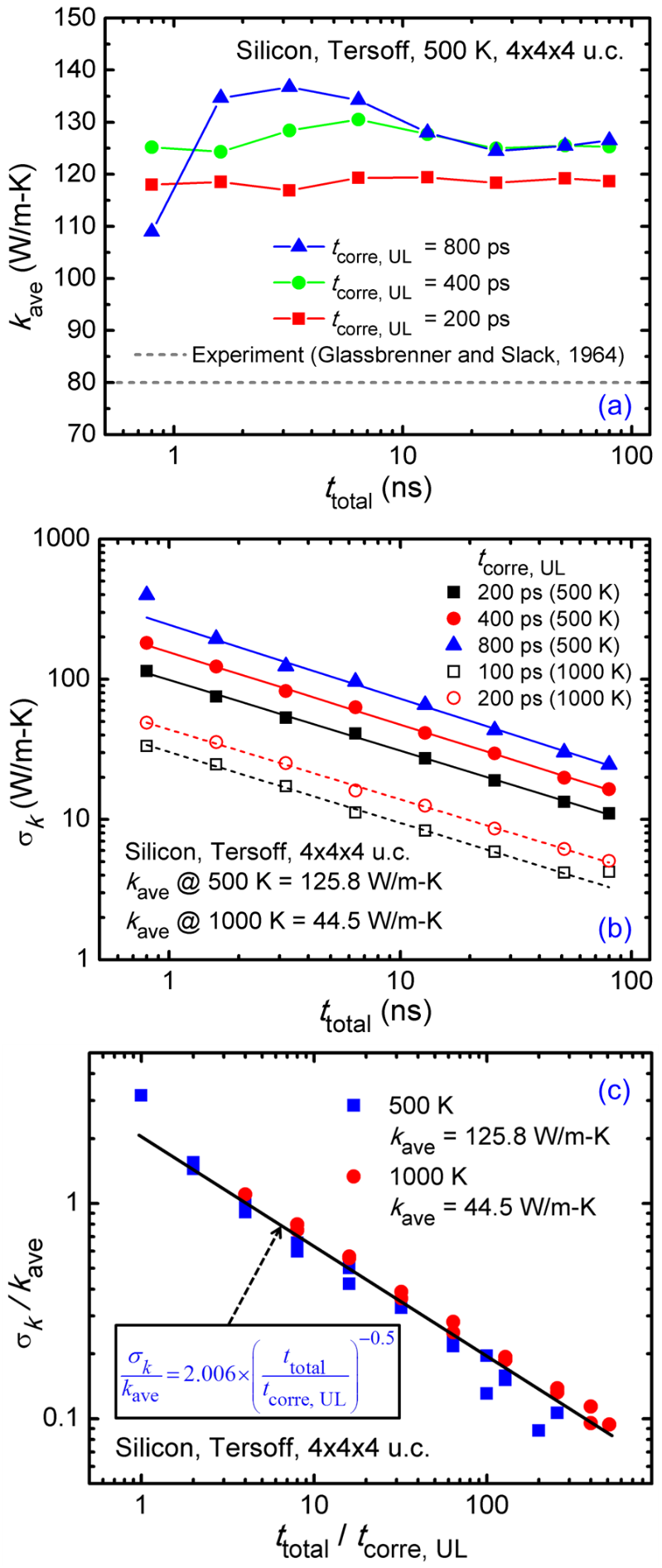} 
\caption{(Color online) (a) Variation of the EMD-predicted average thermal conductivity of silicon with $t_{\text{total}}$ corresponding to a domain size of $4\times4\times4$~u.c. at 500~K. The experimental value (80 W/m-K) by Glassbrenner and Slack~\cite{Glassbrenner1964} is shown as a comparison. (b) Variation of the uncertainty of the thermal conductivity of silicon with $t_{\text{total}}$. (c) Variation of the relative uncertainty of the thermal conductivity of silicon with $t_{\text{total}}/t_{\text{corre, UL}}$. The results in (b) and (c) correspond to a silicon material system with a domain size of $4\times4\times4$~u.c. at 500 and 1000~K. The formula in (c) represents a fit of the data with a square-root relation. }
\label{fig:Fig7_Si_tcorre_ttotal_effects_on_kave_and_dk}
\end{figure}

We move on to quantitatively understand the effects of $t_{\text{total}}$ and $t_{\text{corre, UL}}$ on uncertainty of the EMD-predicted thermal conductivity of silicon. 
In Fig.~\ref{fig:Fig7_Si_tcorre_ttotal_effects_on_kave_and_dk}(a), we show the variation of the average thermal conductivities with $t_{\text{total}}$. For all $t_{\text{corre, UL}}$ values, $k_{\text{ave}}$ converges when $t_{\text{total}}$ is longer than around 10~ns. The converged $k_{\text{ave}}$ generally increases with the increasing $t_{\text{corre, UL}}$, but changes negligibly after $t_{\text{corre, UL}} = 400$~ps, which confirms the appropriateness of the choice of $t_{\text{corre, UL}}$ for the results in Fig.~\ref{fig:Fig6_Si_500K_4x4x4uc_ttotal_effect_on_HCACF_k_distribution}. The much larger EMD-predicted thermal conductivities than the experimental value (80~W/m-K) by Glassbrenner and Slack~\cite{Glassbrenner1964} could be attributed to the Tersoff potential or the defects in the samples used in the experiments. In Fig.~\ref{fig:Fig7_Si_tcorre_ttotal_effects_on_kave_and_dk}(b), we show the variation of the standard deviation of the thermal conductivities with $t_{\text{total}}$, which also includes some results for silicon at 1000~K. For all $t_{\text{corre, UL}}$ values, $\sigma_k$ decreases as $t_{\text{total}}$ increases. We fit the data with a power law relation as $\sigma_k \sim t_{\text{corre, UL}}^{-\gamma}$, and the resulted $\gamma$ varied from 0.48 to 0.52. Similar to the solid argon results, we plotted $\sigma_k/k_{\text{ave}}$ as a function of $t_{\text{total}}/t_{\text{corre, UL}}$, as shown in Fig.~\ref{fig:Fig7_Si_tcorre_ttotal_effects_on_kave_and_dk}(c). By fitting the data with a square-root relation, we obtained $\sigma_k/k_{\text{ave}} = 2.006(t_{\text{total}}/t_{\text{corre, UL}})^{-0.5}$, which is in remarkable agreement with the result for solid argon.

%%%
%%% RESULTS --- Germanium
%%%
\subsection{\label{sec:results-germanium}Germanium}
We also conducted EMD simulations of germanium with a domain size of $4\times4\times4$~u.c. at 500~K. The results are similar to those for silicon. The fitted square-root relation turned out to be $\sigma_k/k_{\text{ave}} = 1.887(t_{\text{total}}/t_{\text{corre, UL}})^{-0.5}$, which is in good agreement with the results for solid argon and silicon. More details about the germanium results can be found in the Supplemental Information.

%%%
%%% RESULTS --- Combined Results and General Consideration
%%%
\subsection{\label{sec:results-combined-results}Combined Results and General Consideration}
\begin{figure}[tbp]
\centering
\includegraphics[width=0.45\textwidth]{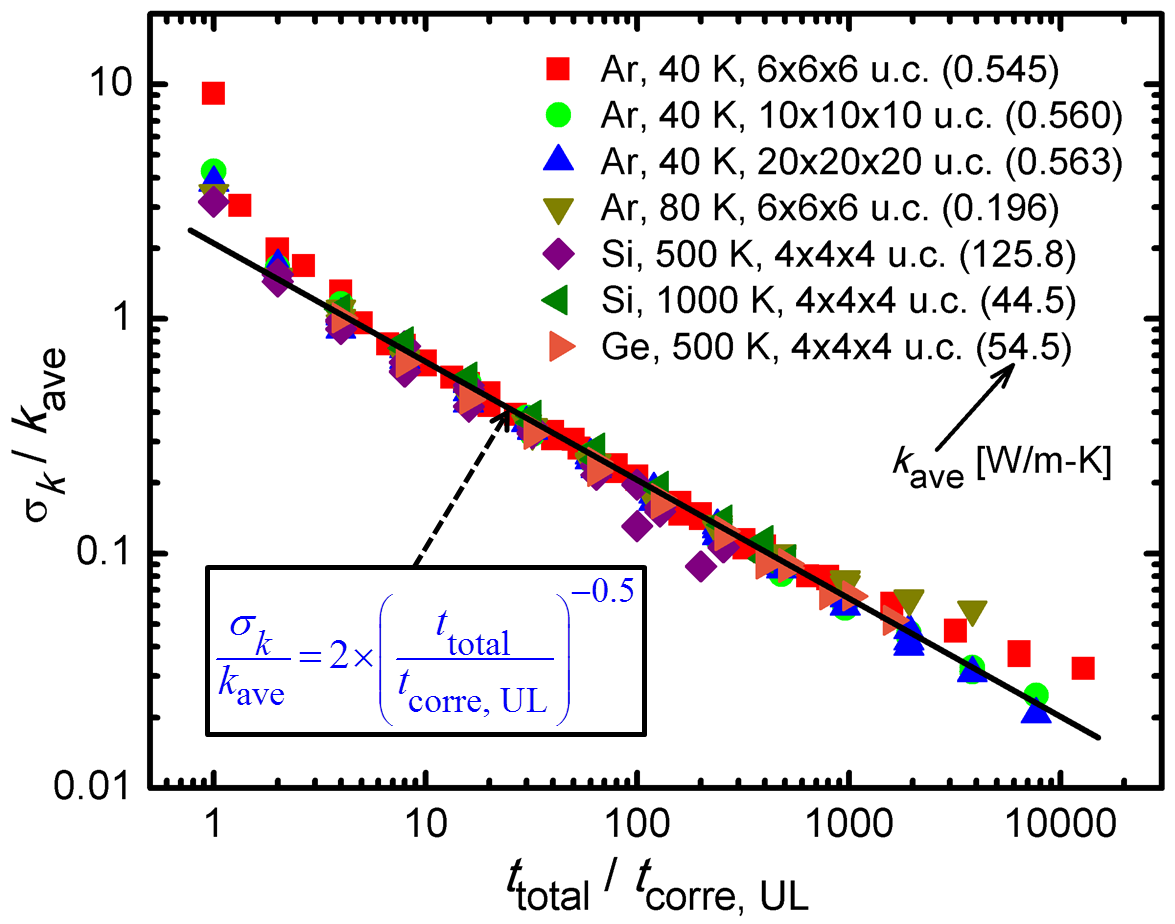} 
\caption{(Color online) Variation of the relative uncertainty of the EMD-predicted thermal conductivity with $t_{\text{corre}}/t_{\text{corre, UL}}$. The data for the different materials and simulation conditions follow a similar trend, which was fit with a single square-root relation. The numbers in the parentheses represent the average thermal conductivities from the corresponding EMD simulations. Note that all the simulation conditions included in this figure have sufficiently long $t_{\text{corre, UL}}$ values, which ensure physically correct thermal conductivity predictions. }
\label{fig:Fig8_dk_over_kave_vs_ttotal_over_tcorre_all}
\end{figure}

Having examined the uncertainty of the EMD-predicted thermal conductivities of solid argon, silicon, and germanium, we provide some discussion on quantifying of the general uncertainty of EMD-predicted thermal conductivities. In Fig.~\ref{fig:Fig8_dk_over_kave_vs_ttotal_over_tcorre_all}, we show the variation of the relative uncertainty of the EMD-predicted thermal conductivities with $t_{\text{total}}/t_{\text{corre, UL}}$, including the results for solid argon, silicon, and germanium under different simulation conditions. It is seen that the data for the different materials and simulation conditions follow a similar trend. When $t_{\text{total}}/t_{\text{corre, UL}}$ is small, the relative uncertainties are large and could be even larger than 100\%. As $t_{\text{total}}/t_{\text{corre, UL}}$ increases, the relative uncertainty decreases. Except for a few data points at extremely small or large $t_{\text{total}}/t_{\text{corre, UL}}$ values, all the data can be fit with a square-root relation, as 
%
%
% eq:sigma_k_relation
\begin{equation}
\frac{\sigma_k}{k_{\text{ave}}} = 2\left(\frac{t_{\text{total}}}{t_{\text{corre, UL}}}\right)^{-0.5}.
\label{eq:sigma_k_relation}
\end{equation}
The independence of the square-root relation of the material system or simulation condition suggests its wide applicability to other material systems or simulation conditions. In other words, this square-root relation could be universal and applicable to all EMD simulations. The previously obtained, different leading constants, 2.025, 2.006, and 1.887 for solid argon, silicon, and germanium, respectively, as compared with the ``2'' in the ``universal'' relation, could be attributed to the limited number of simulations, $t_{\text{total}}$, $t_{\text{corre, UL}}$, and domain size. Note that all the simulation conditions included in Fig.~\ref{fig:Fig8_dk_over_kave_vs_ttotal_over_tcorre_all} have sufficiently long $t_{\text{corre, UL}}$ values, which ensure physically correct thermal conductivity predictions. We point out that according to Sec.~\ref{sec:methodology}, the $\frac{\sigma_k}{k_{\text{ave}}}$ in Eq.~(\ref{eq:sigma_k_relation}) should essentially be understood as $\frac{\sigma_{k_x}}{k_{x, \text{ave}}}$. A derivation of Eq.~(\ref{eq:sigma_k_relation}) is available in the Supplemental Information.
%
%
% fig:Fig8_dk_over_kave_vs_ttotal_over_tcorre_all
%
%

%%%
%%% RESULTS --- How to choose $t_{\text{corre, UL}}$, $t_{\text{total}}$, and $N$ for EMD?
%%%
\subsection{\label{sec:how-to-run-EMD}How to choose $t_{\text{corre, UL}}$, $t_{\text{total}}$, and $N$ for EMD?}
After understanding the uncertainty of EMD-predicted thermal conductivities, we consider a practical question. That is, how should the $t_{\text{corre, UL}}$ and $t_{\text{total}}$ for EMD simulations be appropriately chosen? Because of the statistical nature of MD simulations, there is no easy answer to this question. Here we approach this question by considering the relative error bound ($Q$) together with the confidence level ($P$). We consider first the choice of $t_{\text{corre, UL}}$ and then the choices of $t_{\text{total}}$ for $N=1$ and $N\ge2$, respectively. 

%%%
%%% RESULTS --- Choice of $t_{\text{corre, UL}}$
%%%
\subsubsection{\label{sec:choice-of-tcorreUL}Choice of $t_{\text{corre, UL}}$}
For EMD simulations to provide physically correct thermal conductivity predictions, it is required that the $t_{\text{corre, UL}}$ be sufficiently long so that the truncation error is negligible. In real practice, $t_{\text{corre, UL}}$ is usually determined by inspection of the HCACF profiles, which, although works in some cases, lacks consistency, because different researchers could have very different choices. Here we provide a guideline for choosing $t_{\text{corre, UL}}$ by considering an effective phonon relaxation time, $\tau_{\text{eff}}$, which can be often obtained in two ways: (1) the single-exponential fitting of a HCACF, and (2) the double-exponential fitting of a HCACF. 

%
%
% table:fit_time_constant
\begin{table*}[htbp]
\caption{Summary of effective phonon relaxation times for solid argon at 40~K, silicon at 500~K, and germanium at 500~K, obtained from the single- and double-exponential fitting.}
\renewcommand{\arraystretch}{1.5}
\begin{tabular}{c c c c c c c c c}
\hline
\hline
\multirow{3}{*}{Material} & & \multicolumn{2}{c}{Single Exp. Fit} & & \multicolumn{4}{c}{Double Exp. Fit} \\
\cline{3-4} \cline{6-9}
                  & & $A_1$ & $\tau_{\text{eff, 1}}$ & & $A_{21}$ & $\tau_{\text{eff, 21}}$ & $A_{22}$ & $\tau_{\text{eff, 22}}$ \\
                  & & (eV$^2$A$^2$/ps$^2$) & (ps) & & (eV$^2$A$^2$/ps$^2$) & (ps) & (eV$^2$A$^2$/ps$^2$) & (ps) \\
\hline
Ar at 40~K & & 0.759 & 1.565 & & 1.029 & 0.274 & 0.332 & 3.769   \\
Si at 500~K & & 811.7 & 16.65 & & 568.3 & 4.650 & 450.9 & 28.45 \\
Ge at 500~K & & 369.8 & 18.54 & & 246.7 & 6.290 & 195.2 & 31.75 \\
\hline
\hline
%\multicolumn{9}{l}{Note: The units for $A$ and $\tau$ are (eV$^2$A$^2$/ps$^2$) and (ps), respectively.}
\end{tabular}
\label{table:fit_time_constant}
\end{table*} 

The single-exponential fitting of a HCACF is based on the concept of gray phonons, \textit{i.e.}, all phonons have the same effective relaxation time. The relevant formula reads, 
%
%
% eq:single_exp_fit
\begin{equation}
\text{HCACF}_1(t_{\text{corre}}) = A_1\exp\left(-\frac{t_{\text{corre}}}{\tau_{\text{eff, 1}}}\right), 
\label{eq:single_exp_fit}
\end{equation}
where $A_1$ and $\tau_{\text{eff, 1}}$ are fitting parameters. Alternatively, $A_1$ could be fixed at HCACF(0), leaving $\tau_{\text{eff, 1}}$ as the only fitting parameter. But we find that this latter method typically results in poorer fitting results as compared to the first method. We used this method to fit the HCACF for solid argon at 40~K, silicon at 500~K, and germanium at 500~K, as seen in Fig.~\ref{fig:Fig3_Ar_40K_6x6x6uc_ttotal_effect_on_HCACF_k_distribution}(d) for solid argon and Fig.~\ref{fig:Fig6_Si_500K_4x4x4uc_ttotal_effect_on_HCACF_k_distribution}(d) for silicon (the germanium result is available in the Supplemental Information). 

The double-exponential fitting of a HCACF is based on the formula,~\cite{McGaughey2004} 
%
%
% eq:double_exp_fit
\begin{align}
\text{HCACF}_2(t_{\text{corre}}) &= A_{21}\exp \left( -\frac{t_{\text{corre}}}{\tau_{\text{eff, 21}}}\right) \nonumber \\ 
&+ A_{22}\exp\left(-\frac{t_{\text{corre}}}{\tau_{\text{eff, 22}}} \right), 
\label{eq:double_exp_fit}
\end{align}
where $A_{21}$, $\tau_{\text{eff, 21}}$, $A_{22}$, and $\tau_{\text{eff, 22}}$ are fitting parameters. This method divides phonons into two broad categories: those with a short relaxation time and those with a long relaxation time. Despite this relatively coarse treatment, this method typically provides much better fitting results than the single-exponential fitting method. We used this method to fit the HCACF for solid argon at 40~K, silicon at 500~K, and germanium at 500~K, as seen in Fig.~\ref{fig:Fig3_Ar_40K_6x6x6uc_ttotal_effect_on_HCACF_k_distribution}(d) for solid argon and Fig.~\ref{fig:Fig6_Si_500K_4x4x4uc_ttotal_effect_on_HCACF_k_distribution}(d) for silicon (the germanium result is available in the Supplemental Information). 

Table~\ref{table:fit_time_constant} summarizes the effective phonon relaxation times obtained from the single- and double-exponential fitting. It is seen that $\tau_{\text{eff, 1}}$ lies between $\tau_{\text{eff, 21}}$ and $\tau_{\text{eff, 22}}$ (roughly at the middle). As a validation, the solid argon results agree well with previous results by McGaughey and Kaviany.~\cite{McGaughey2004} Considering the higher accuracy of the double-exponential fitting, we recommend using $\tau_{\text{eff, 22}}$ (the longer time constant) as the effective phonon relaxation time to determine the $t_{\text{corre, UL}}$. Hereafter, we will refer to this time constant as $\tau_{\text{eff}}$ for simplicity. From Eq.~(\ref{eq:double_exp_fit}), we can obtain the relative truncation error due to a finite $t_{\text{corre, UL}}$, as
%
%
% eq:double_exp_fit_e_truncation
\begin{equation}
E_{\text{trunc}} (t_{\text{corre, UL}}) = \frac
{\int_{t_{\text{corre, UL}}}^{\infty} \! \text{HCACF}_2(t_{\text{corre}}) \; \mathrm{d} t_{\text{corre}}}
{\int_0^{\infty} \! \text{HCACF}_2(t_{\text{corre}}) \; \mathrm{d} t_{\text{corre}}}.
\label{eq:double_exp_fit_e_truncation}
\end{equation}
Substituting Eq.~(\ref{eq:double_exp_fit}) into Eq.~(\ref{eq:double_exp_fit_e_truncation}) and carrying out the integrations, we obtain
%
%
% eq:double_exp_fit_e_truncation2
\begin{align}
E_{\text{trunc}} (t_{\text{corre, UL}}) &= \frac
{A_{21}\tau_{\text{eff, 21}} \exp \left( -\frac{t_{\text{corre, UL}}}{\tau_{\text{eff, 21}}} \right)}
{A_{21}\tau_{\text{eff, 21}} + A_{22}\tau_{\text{eff, 22}}} \nonumber \\
&+\frac
{A_{22}\tau_{\text{eff, 22}} \exp \left( -\frac{t_{\text{corre, UL}}}{\tau_{\text{eff, 22}}} \right)}
{A_{21}\tau_{\text{eff, 21}} + A_{22}\tau_{\text{eff, 22}}}.
\label{eq:double_exp_fit_e_truncation2}
\end{align}
Considering $\tau_{\text{eff, 21}} \ll \tau_{\text{eff, 22}}$ and $A_{21}\tau_{\text{eff, 21}} \ll A_{22}\tau_{\text{eff, 22}}$, which are typically true, we have
%
%
% eq:double_exp_fit_e_truncation3
\begin{equation}
E_{\text{trunc}} (t_{\text{corre, UL}}) \approx \exp \left( -\frac{t_{\text{corre, UL}}}{\tau_{\text{eff, 22}}} \right).
\label{eq:double_exp_fit_e_truncation3}
\end{equation}
As a result, we recommend choosing $t_{\text{corre, UL}}$ to be $(5\sim10)\tau_{\text{eff}}$ to achieve an $E_{\text{trunc}}$ smaller than 1\% (Note that $\exp(-5) \approx 0.67\%$). In this study, most of the $t_{\text{corre, UL}}$ values were conservatively chosen to be larger than 10$\tau_{\text{eff}}$. Note that the double-exponential fitting only needs to be done once. After $t_{\text{corre, UL}}$ is determined, the additional EMD simulations can be done by using the direct integration method to calculate the thermal conductivity.

%%%
%%% RESULTS --- Choice of $t_{\text{total}}$ for $N=1$
%%%
\subsubsection{\label{sec:choice-of-ttotal-N=1}Choice of $t_{\text{total}}$ with $N=1$}
Here we consider a scenario, in which only one EMD simulation is conducted to calculate the thermal conductivity. From the results shown in Figs.~\ref{fig:Fig2_Ar_k_distribution_velocity_seed_effect} and \ref{fig:Fig4_Ar_40K_tcorre_total_effects_on_kave_and_dk}, we realize that the thermal conductivities from independent simulations can be reasonably assumed to follow a normal distribution. As a result, the predicted thermal conductivity in a particular direction (say $k_x$, without loss of generality) from just one simulation will fall in the range between ($k_{x, \text{ave}} - \sigma_{k_x}$) and ($k_{x, \text{ave}} + \sigma_{k_x}$) with a confidence level of 68.3\%. 

If we define the relative error of an EMD-predicted thermal conductivity as $\frac{|k_{x, 1} - k_{x, \text{ave}}|}{k_{x, \text{ave}}}$, where $k_{x, 1}$ is the thermal conductivity from a single EMD simulation and $k_{x, \text{ave}}$ is the average thermal conductivity from a large number of EMD simulations (or the ``true'' thermal conductivity), and the bound of the relative error as $Q$, then the confidence level, $P$, corresponding to $Q$ can be expressed as
%
%
% eq:k_EMD_confidence
\begin{equation}
P = P\left(\frac{|k_{x, 1} - k_{x, \text{ave}}|}{k_{x, \text{ave}}} \le Q\right), 
\label{eq:k_EMD_confidence}
\end{equation}
or,
%
%
% eq:k_EMD_confidence2
\begin{equation}
P = P\left[k_{x, \text{ave}} (1-Q) \le k_{x, 1} \le k_{x, \text{ave}} (1+Q)\right]. 
\label{eq:k_EMD_confidence2}
\end{equation}
Considering the cumulative distribution function (CDF) of a normal distribution, $\mathcal{N}(\mu, \sigma^2)$ (with a mean $\mu$ and a standard deviation $\sigma$),\cite{Rice2007}  
%
%
% eq:normaldistcdf
\begin{equation}
\text{CDF}(x) = \frac{1}{2}\left[ 1 +  \text{erf}\left( \frac{x-\mu}{\sigma\sqrt{2}} \right) \right], 
\label{eq:normaldistcdf}
\end{equation}
we have 
%
%
% eq:k_EMD_confidence3
\begin{equation}
P = \text{CDF}\left[ k_{x, \text{ave}} (1+Q) \right] - \text{CDF}\left[ k_{x, \text{ave}} (1-Q) \right], 
\label{eq:k_EMD_confidence3}
\end{equation}
which can be simplified by applying Eq.~(\ref{eq:normaldistcdf}) and the property that $\text{erf}(-x) = -\text{erf}(x)$, into
%
%
% eq:k_EMD_confidence4
\begin{equation}
P = \text{erf}\left( \frac{k_{x, \text{ave}}Q}{\sigma_{k_x}\sqrt{2}} \right). 
\label{eq:k_EMD_confidence4}
\end{equation}
Finally, we can substitute Eq.~(\ref{eq:sigma_k_relation}) into Eq.~(\ref{eq:k_EMD_confidence4}) to obtain
%
%
% eq:k_EMD_confidence5
\begin{equation}
P = \text{erf}\left(\sqrt{\frac{t_{\text{total}}}{8t_{\text{corre, UL}}}}Q\right), 
\label{eq:k_EMD_confidence5}
\end{equation}
which can alternatively be re-arranged as
%
%
% eq:no_of_EMD_simulations
\begin{equation}
\frac{t_{\text{total}}}{t_{\text{corre, UL}}} = 8\left[ \frac{\text{erf}^{-1}(P)}{Q} \right]^2. 
\label{eq:no_of_EMD_simulations}
\end{equation}
Here the ``erf and ``erf$^{-1}$'' stand for the error function and inverse error function, respectively. Therefore, the choice of $t_{\text{total}}$ directly depends on the desired relative error bound and confidence level. 
%
%
% fig:Fig9_Determine_no_of_EMD_simulations
\begin{figure}[tbp]
\centering
\includegraphics[width=0.45\textwidth]{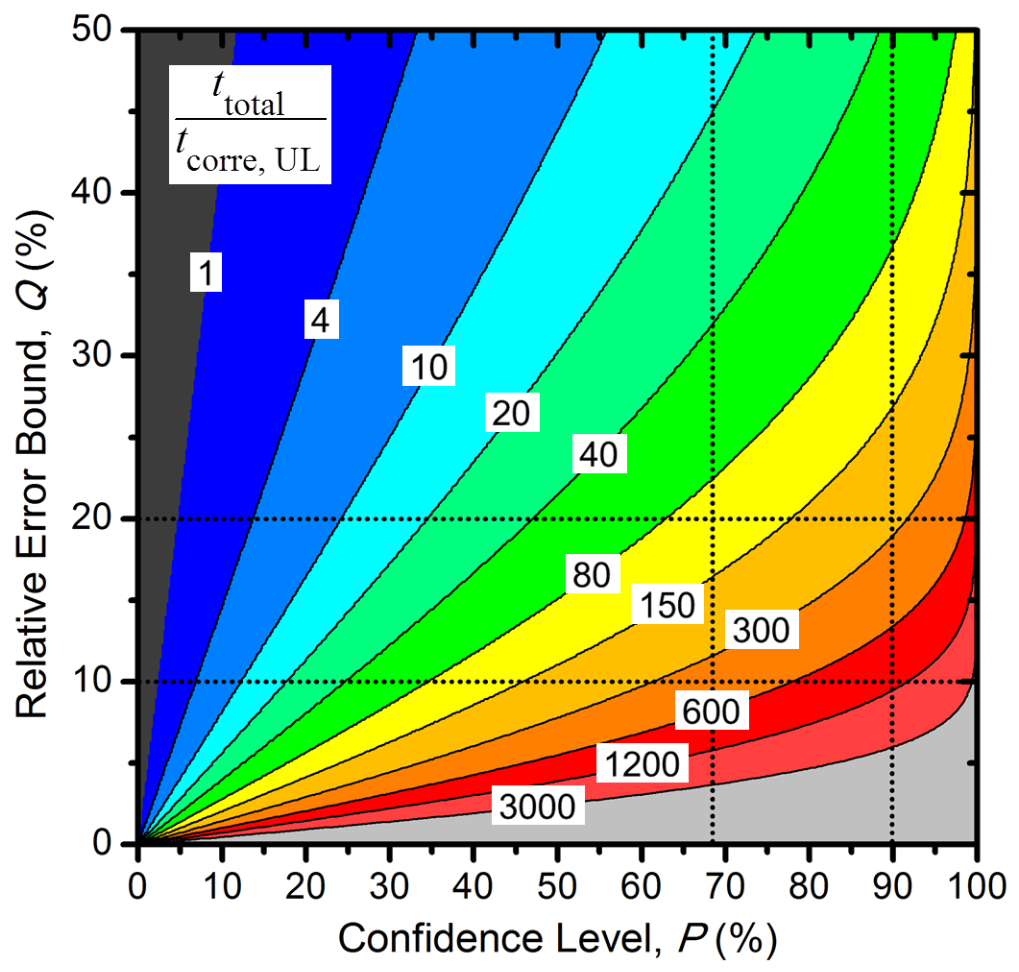} 
\caption{(Color online) Contour plot $t_{\text{total}}/t_{\text{corre, UL}}$ with the relative error bound ($Q$) and confidence level ($P$) for EMD simulations. This figure provides a guideline for choosing $t_{\text{total}}$ (the choice of $t_{\text{corre, UL}}$ is discussed in Sec.~\ref{sec:choice-of-tcorreUL}) to achieve a desired relative error bound with a desired confidence level. The $t_{\text{total}}$ should be replaced by $N\times t_{\text{total}}$, if $N$ independent simulations are conducted, and by $3N\times t_{\text{total}}$, if the material can be further assumed to be isotropic. The horizontal dotted lines indicate relative error bounds of 10\% and 20\%. The vertical dotted lines indicate confidence levels of 68.3\% and 90\%. } 
\label{fig:Fig9_Determine_no_of_EMD_simulations}
\end{figure} 

In Fig.~\ref{fig:Fig9_Determine_no_of_EMD_simulations}, we show a contour plot of $\frac{t_{\text{total}}}{t_{\text{corre, UL}}}$ as a function of $P$ and $Q$. We focus on relative error bounds in the range from 0\% to 50\% and confidence levels from 0\% to 100\%. It is seen that at a constant $Q$, $\frac{t_{\text{total}}}{t_{\text{corre, UL}}}$ increases with the increasing $P$, whereas at a constant $P$, $\frac{t_{\text{total}}}{t_{\text{corre, UL}}}$ decreases with the increasing $Q$. Note that only the results with $\frac{t_{\text{total}}}{t_{\text{corre, UL}}} \ge 1$ are physically correct, because $t_{\text{total}} \ge t_{\text{corre, UL}}$. In the following we provide two specific examples to illustrate the use of Eq.~(\ref{eq:no_of_EMD_simulations}). (1) Consider $k_x$ of solid argon at 40~K, if we target at $Q = 10\%$ and $P = 90\%$, and choose $t_{\text{corre, UL}} = 40$~ps (around $10.6\tau_{\text{eff}}$), then $t_{\text{total}} = 43.3$~ns. (2) Consider $k_x$ of silicon at 500~K, if we target at $Q = 20\%$ and $P = 68.3\%$, and choose $t_{\text{corre, UL}} = 400$~ps (around $14.1\tau_{\text{eff}}$), then $t_{\text{total}} = 40.1$~ns. We notice that the $t_{\text{total}}$ values are typically very large in order to achieve a relatively small relative error bound with a relatively high confidence level. The large $t_{\text{total}}$ values, however, could be reduced by conducting multiple simulations, as discussed in the following section.

%%%
%%% RESULTS --- Choice of $t_{\text{total}}$ for $N\ge2$
%%%
\subsubsection{\label{sec:choice-of-ttotal-N>=2}Choice of $t_{\text{total}}$ with $N\ge2$}
\begin{figure}[tbp]
\centering
\includegraphics[width=0.45\textwidth]{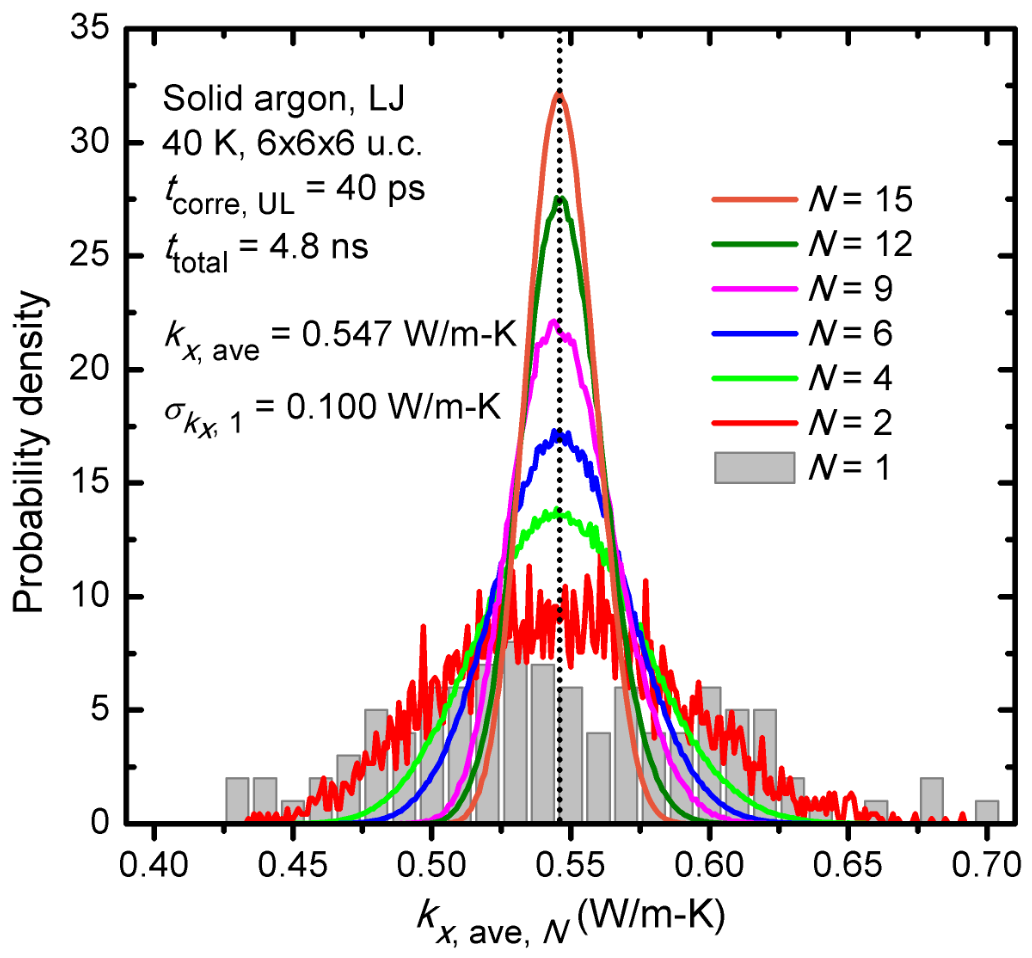} 
\caption{(Color online) Distribution of the average thermal conductivity of solid argon from $N$ independent simulations. The simulation conditions are indicated in the figure. The distributions corresponding to different $N$ values all qualitatively follow a normal distribution. The mean of the distribution remains the same (0.547~W/m-K), but the standard deviation decreases, as $N$ increases. } 
\label{fig:Fig10_NoofEMD_kProfile}
\end{figure}

It is a typical practice that EMD simulations are conducted with multiple independent runs, and the average thermal conductivity is calculated. According to the Central Limit Theorem,~\cite{Rice2007} if $N$ independent simulations are conducted, then the standard deviation of the average thermal conductivity distribution will decrease from $\sigma_{k_x}$ to $\frac{\sigma_{k_x}}{\sqrt{N}}$. To illustrate this fact, we consider the distribution of the average thermal conductivity of solid argon calculated from $N$ independent simulations (\textit{i.e.}, $k_{x, \text{ave}, N}$), as shown in Fig.~\ref{fig:Fig10_NoofEMD_kProfile}. The simulations correspond to solid argon at 40~K with a domain size of $6\times6\times6$~u.c., $t_{\text{corre, UL}} = 40$~ps, and $t_{\text{total}} = 4.8$~ns. We varied $N$ from 1 to 15. It is seen that the distributions corresponding to different $N$ values all qualitatively follow a normal distribution. We also observe that the mean of the distribution remains the same (0.547~W/m-K), but the standard deviation decreases, as $N$ increases. 
%
%
% fig:Fig10_NoofEMD_kProfile

Substituting $\sigma_{k_x}$ with $\frac{\sigma_{k_x}}{\sqrt{N}}$ and repeating the derivations in Sec.~\ref{sec:choice-of-ttotal-N=1}, we obtain
%
%
% eq:no_of_EMD_simulations2
\begin{equation}
\frac{N\times t_{\text{total}}}{t_{\text{corre, UL}}} = 8\left[ \frac{\text{erf}^{-1}(P)}{Q} \right]^2. 
\label{eq:no_of_EMD_simulations2}
\end{equation}
Consider again the two examples given in Sec.~\ref{sec:choice-of-ttotal-N=1}. We realize that the $t_{\text{total}}$ for the first and second examples can be reduced to 4.33 and 4.01~ns, respectively, by conducting 10 independent simulations (\textit{i.e.}, $N=10$). Equation~(\ref{eq:no_of_EMD_simulations2}) can also be used to determine $N$, if $t_{\text{total}}$ is restricted upfront. It is worth mentioning that Equation~(\ref{eq:no_of_EMD_simulations}) provides a mathematical demonstration of the equivalence of time-averaging and ensemble-averaging. At a fixed $t_{\text{corre, UL}}$, a desired $Q$ and $P$ can be achieved by maintaining ($N\times t_{\text{total}}$) to be a constant, which can be realized by running either a small number of long simulations or a large number of short simulations. 

For isotropic materials, Equation~(\ref{eq:no_of_EMD_simulations2}) can be further written as
%
%
% eq:no_of_EMD_simulations3
\begin{equation}
\frac{3N\times t_{\text{total}}}{t_{\text{corre, UL}}} = 8\left[ \frac{\text{erf}^{-1}(P)}{Q} \right]^2, 
\label{eq:no_of_EMD_simulations3}
\end{equation}
where the ``3'' accounts for the $x$, $y$, and $z$ directions. Therefore, taking into account the isotropicity of materials could further reduce $t_{\text{total}}$ by 66.7\% . 

We provide a validation for Eq.~(\ref{eq:no_of_EMD_simulations3}) by using the data for solid argon at 40~K with a domain size of $6\times6\times6$~u.c., $t_{\text{corre, UL}} = 40$~ps, $t_{\text{total}} = 4.8$~ns, and $N = 6$, which are shown in Fig.~\ref{fig:Fig10_NoofEMD_kProfile}. By considering the combinations of 6 simulations out of 100 simulations ($^{100}C_6 = 1,192,052,400$), we find that the confidence level for a relative error bound of 10\% (\textit{i.e.}, $0.9k_{x, \text{ave}} \le k_{x, \text{ave}, 6} \le 1.1k_{x, \text{ave}}$, or $0.492 \le k_{x, \text{ave}, 6} \le 0.602$) is 98.5\%. On the other hand, from Eq.~(\ref{eq:no_of_EMD_simulations3}), we have $P = \text{erf}\left( \sqrt{\frac{3\times6\times4.8}{8\times0.04}}\times10\% \right)$ = 98.0\%. The excellent agreement of these two confidence levels provides a validation for Eq.~(\ref{eq:no_of_EMD_simulations3}) (or the original Eq.~(\ref{eq:no_of_EMD_simulations})). The slight discrepancy could be attributed to the limited total number of simulations (\textit{i.e.}, 100).

%%%
%%% RESULTS --- How to report EMD-predicted $k_x$?
%%%
\subsection{\label{sec:how-to-report-EMD}How to report EMD-predicted $k_x$?}
\begin{figure*}[tbp]
\centering
\includegraphics[width=0.9\textwidth]{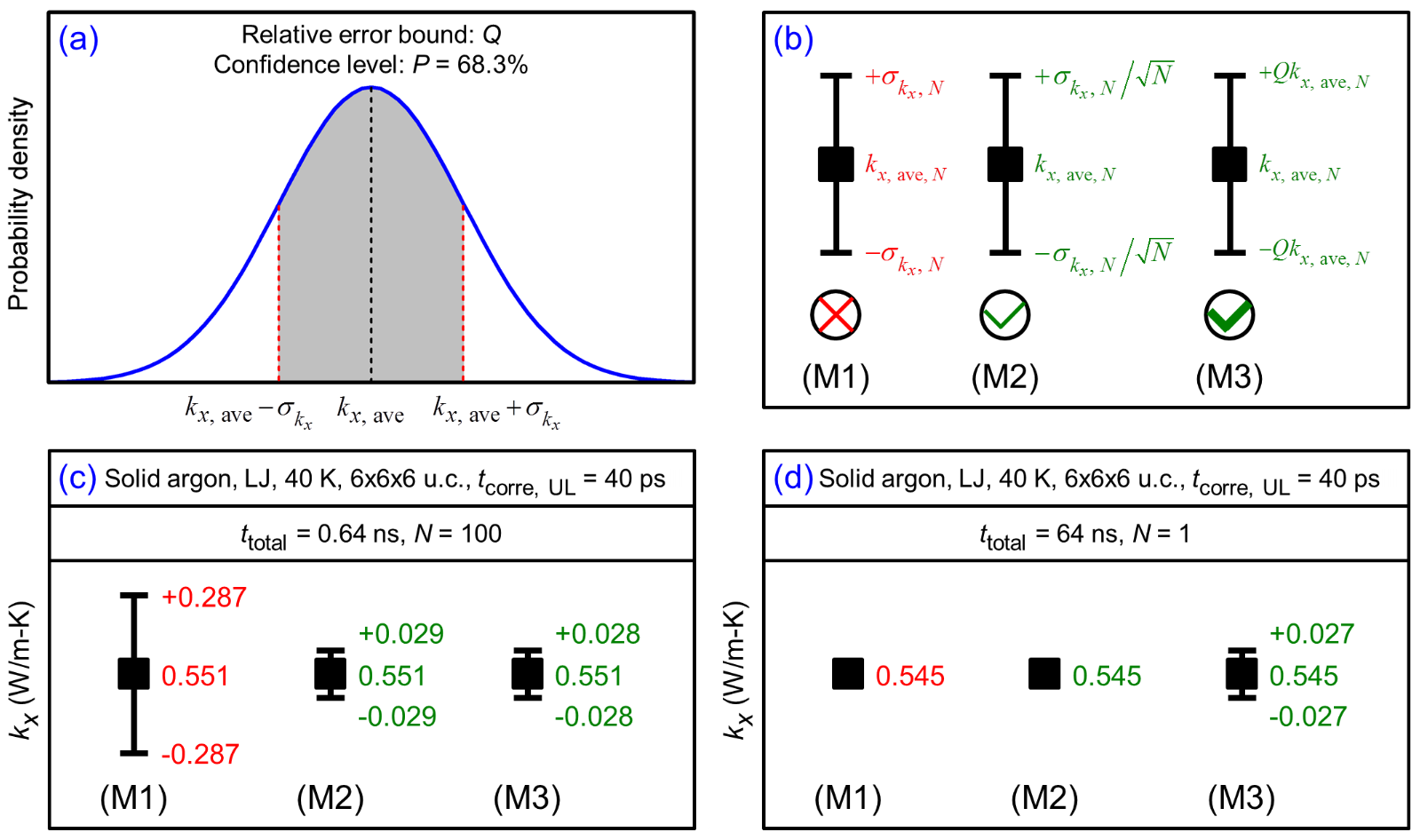} 
\caption{(Color online) (a) Illustration of a normal distribution with a mean (or average), $k_{x, \text{ave}}$, and a standard deviation, $\sigma_{k_x}$. (2) Illustration of three possible ways to report $k_x$ from EMD simulations. (3) An example of reporting $k_x$ from 100 EMD simulations with a short $t_{\text{total}}$. (4) An example of reporting $k_x$ from one EMD simulation with a long $t_{\text{total}}$. The error bars are not drawn to scale. }
\label{fig:Fig11_HowToReportData}
\end{figure*}

Having understood the uncertainty of EMD simulations, we consider how EMD-predicted thermal conductivities should be reported. In Fig.~\ref{fig:Fig11_HowToReportData}(a), we show a normal distribution with a mean (or average), $k_{x, \text{ave}}$, and a standard deviation, $\sigma_{k_x}$. If one simulation is conducted, then the $k_x$ will have a confidence level of 68.3\% to fall between $k_{x, \text{ave}} - \sigma_{k_x}$ and $k_{x, \text{ave}} + \sigma_{k_x}$. We note that typically a confidence level of 68.3\% is assumed by default in the literature when EMD-predicted thermal conductivities are reported with error bars. In Fig.~\ref{fig:Fig11_HowToReportData}(b), we show three possible methods of reporting EMD-predicted $k_x$: (M1) using $k_{x, \text{ave}, N}$ and $\sigma_{k_x, N}$, (M2) using $k_{x, \text{ave}, N}$ and $\sigma_{k_x, N}/\sqrt{N}$ (named the ``standard error''), and (M3) using $k_{x, \text{ave}, N}$ and $\sigma_{k_x, N}$. In all the three methods, the average value of the $N$ EMD simulations, $k_{x, \text{ave}, N}$ is considered the best estimate of the ``true'' thermal conductivity, $k_{x, \text{ave}}$, but they report the error bar in different ways. In (M1), the error bar emphasizes the distribution of the $k_x$ values from the $N$ simulations (\textit{i.e.}, precision), instead of how close the predicted $k_{x, \text{ave}, N}$ is to the $k_{x, \text{ave}}$ (\textit{i.e.}, accuracy). A larger $N$ could even counter-intuitively result in a larger error bar. As a result, we consider this method to be inappropriate for reporting EMD-predicted $k_x$, despite the fact that it has been widely used in previously studies. In (M2), the error bar emphasizes the distribution of $k_{x, \text{ave}, N}$ (rather than $k_{x}$) and thus the accuracy of the predictions. It also correctly reflects the fact that a larger $N$ will result in a smaller error bar. As a result, we consider this method to be appropriate for reporting EMD-predicted $k_x$. In (M3), the error bar is evaluated by using Eq.~(\ref{eq:sigma_k_relation}), where the $t_{\text{total}}$ should be replaced by $N\times t_{\text{total}}$ if $N$ simulations are conducted. Similar to (M2), (M3) emphasizes the accuracy of the predictions. We consider this method to be appropriate for reporting EMD-predicted $k_x$, because Eq.~(\ref{eq:sigma_k_relation}) is obtained from statistical analysis of a large number of data, and (M3) also incorporates the equivalence of time-averaging and assemble-averaging. Under the assumption that $k_{x, \text{ave}, N} \approx k_{x, \text{ave}}$ (or $Q\approx0$), (M2) and (M3) can be shown to be equivalent by using Eq.~(\ref{eq:sigma_k_relation}). To illustrate how these three methods work, we consider EMD-predicted $k_x$ of solid argon at 40~K with a domain size of $6\times6\times6$~u.c. and $t_{\text{corre, UL}} = 40$~ps, as shown in Fig.~\ref{fig:Fig11_HowToReportData}(c) and (d), which correspond to ($t_{\text{total}} = 0.64$~ns, $N = 100$) and ($t_{\text{total}} = 64$~ns, $N = 1$), respectively. From Fig.~\ref{fig:Fig11_HowToReportData}(c), it is seen that (M1) results in a large error bar, even though 100 simulations are conducted, whereas (M2) and (M3) results in small and nearly equally-sized error bars. From Fig.~\ref{fig:Fig11_HowToReportData}(d), it is seen that no error bar is predicted by (M1) and (M2) since only one simulation is conducted, but (M3) predicts an error bar, which is nearly the same with the (M3)-predicted error bar in Fig.~\ref{fig:Fig11_HowToReportData}(c). According to the equivalence of time-averaging and assemble-averaging, the error bar for ($t_{\text{total}} = 0.64$~ns, $N = 100$) and ($t_{\text{total}} = 64$~ns, $N = 1$) should be same. Since (M3) correctly predicts this equivalence, we consider it to be the most appropriate method among the three for reporting EMD-predicted $k_x$. We recommend using (M3) for future EMD studies. 
%
%
% fig:Fig11_HowToReportData

In the following we provide a summary of the key steps of doing EMD simulations.

(1) Choose a $t_{\text{corre, UL}}$ according to the guideline in Sec.~\ref{sec:choice-of-tcorreUL}. 

(2) Choose a desired $Q$ and $P$ (typically 68.3\%). 

(3) Choose a $t_{\text{total}}$ according to the guideline in Sec.~\ref{sec:choice-of-ttotal-N=1}, or a $t_{\text{total}}$ and an $N$ according to the guideline in Sec.~\ref{sec:choice-of-ttotal-N>=2}. 

(4) Conduct the simulations.

(5) Report the EMD-predicted thermal conductivities according to the guideline in Sec.~\ref{sec:how-to-report-EMD}.

A few remarks: (1) A $t_{\text{total}}$ or a $t_{\text{total}}$ and an $N$ can also be determined first, and then the $Q$ calculated with Eq.~(\ref{eq:no_of_EMD_simulations}), (\ref{eq:no_of_EMD_simulations2}), or (\ref{eq:no_of_EMD_simulations3}) under a certain $P$ (typically 68.3\%). (2) The time step of the EMD simulations should be chosen appropriately to capture the physics of the phonon transport at an acceptable computational cost. In the literature, it is typically chosen as around 1/50th of the period of the highest frequency phonon, which can be determined from lattice dynamics calculations, for example, by using \texttt{GULP}.~\cite{Gale2003} We expect that the time step could be chosen as large as around 1/10th of the period of the highest frequency phonon; further studies could be conducted on this topic. (3) A size effect study should be conducted before meaningful thermal conductivity data are collected. 

Finally, we point out that although this study focuses on EMD-predicted thermal conductivities of isotropic materials, the key findings (\textit{e.g.}, Eqs.~(\ref{eq:sigma_k_relation}) and (\ref{eq:no_of_EMD_simulations})) should also apply to other properties calculated with the Green-Kubo formula or its counterpart(s) (\textit{e.g.}, interfacial thermal conductance from EMD simulations~\cite{Chalopin2012}), or anisotropic materials (\textit{e.g.}, graphite, bismuth telluride). In addition, we have done some sensitivity analysis to evaluate the relative effects of the velocity initialization seed, $t_{\text{corre, UL}}$, and $t_{\text{total}}$ on the thermal conductivity predictions, as shown in the Supplemental Information.

%%%
%%% CONCUSIONS
%%%
\section{\label{sec:conclusions}Conclusions}
In summary, this paper provides a study on quantifying the uncertainty of the EMD-predicted thermal conductivities by using solid argon, silicon, and germanium as model materials systems. We find that the uncertainty increases with the upper limit of the correlation time, $t_{\text{corre, UL}}$, and decreases with the total simulation time, $t_{\text{total}}$, whereas the velocity initialization seed, simulation domain size, temperature, and type of material have minimal effects. We have obtained a ``universal'' square-root relation for quantifying the uncertainty, as $\sigma_{k_x}/k_{x, \text{ave}} = 2(t_{\text{total}}/t_{\text{corre, UL}})^{-0.5}$. With this relation, it is possible to predict the uncertainty of the thermal conductivities from EMD simulations based on the chosen simulation parameters, even before the simulations are done. We have also conducted statistical analysis of the EMD-predicted thermal conductivities and derived a formula that correlates the relative error bound ($Q$), confidence level ($P$), $t_{\text{corre, UL}}$, $t_{\text{total}}$, and number of independent simulations ($N$). We recommend choosing $t_{\text{corre, UL}}$ to be 5--10 times the effective phonon relaxation time, $\tau_{\text{eff}}$, and choosing $t_{\text{total}}$ and $N$ based on the desired relative error bound and confidence level. We also recommend reporting EMD-predicted thermal conductivities as $k_{x, \text{ave}, N} (1 \pm Q)$, with the confidence level indicated. This study provides new insights into understanding the uncertainty of EMD-predicted thermal conductivities. It also provides a guideline for running EMD simulations to achieve a desired relative error bound with a desired confidence level and for reporting EMD-predicted thermal conductivities.

%%%
%%% ACKNOWLEDGMENTS
%%%
\section*{\label{sec:acknowledgments}Acknowledgments}
Z. Wang and X. Ruan would like to thank the support from the National Science Foundation (Award No. 1150948). S. Safarkhani and G. Lin would like to acknowledge the support by the U.S. Department of Energy, Office of Science, Office of Advanced Scientific Computing Research, Applied Mathematics program as part of the Multifaceted Mathematics for Complex Energy Systems (M$^2$ACS) project and part of the Collaboratory on Mathematics for Mesoscopic Modeling of Materials project, and NSF Grant DMS-1555072.

%%%
%%% REFERENCES
%%%
%\section{\label{sec:references}References}


\begin{thebibliography}{99}
% [1]
% Green-Kubo method
\bibitem{Green1954}
M. S. Green, 
\href{http://scitation.aip.org/content/aip/journal/jcp/22/3/10.1063/1.1740082}
	{J. Chem. Phys. \textbf{22}, 398 (1954)}.

% Green-Kubo method, fluctuation-dissipation theorem
\bibitem{Kubo1957}
R. Kubo, 
\href{http://journals.jps.jp/doi/abs/10.1143/JPSJ.12.570}
	{J. Phys. Soc. Jpn. \textbf{12}, 570 (1957)}.

% EMD and NEMD on silicon, size effect, 
% Green-Kubo formula, heat current formula, direct method
\bibitem{Schelling2002}
P. K. Schelling, S. R. Phillpot, and P. Keblinski, 
\href{http://journals.aps.org/prb/abstract/10.1103/PhysRevB.65.144306}
	{Phys. Rev. B \textbf{65}, 144306 (2002)}.

% Green-Kubo method
\bibitem{McQuarrie2000}
D. A. McQuarrie, 
\textit{Statistical Mechanics} 
(University Science Books, Sausalito, 2000).

% argon material system, EMD, thermal conductivity of solid argon
% double exponential fitting of HCACF
\bibitem{McGaughey2004}
A. J. H. McGaughey and M. Kaviany, 
\href{http://www-personal.umich.edu/~kaviany/researchtopics/ijhmt1.pdf}
	{Int. J. Heat Mass Transfer \textbf{47} (8), 1783 (2004)}.

% [6]
% SED method, freestanding and supported silicene
\bibitem{Wang2015}
Z. Wang, T. Feng, and X. Ruan, 
\href{http://scitation.aip.org/content/aip/journal/jap/117/8/10.1063/1.4913600}
	{J. Appl. Phys. \textbf{117}, 084317 (2015)}.

% EMD, solid argon
\bibitem{Wang2016a}
Z. Wang and X. Ruan, 
\href{http://www.sciencedirect.com/science/article/pii/S092702561630177X}
	{Comput. Mater. Sci. \textbf{121}, 97 (2016)}.

% EMD, uncertainty quantification, IMECE paper
\bibitem{Wang2016b}
Z. Wang and X. Ruan, 
\href{http://zuyuanwang.net/Publications/Conference/2016_Wang_EMD_Uncertainty_IMECE2016-68083.pdf} 
	{IMECE2016-68083},
	Proceedings of the ASME 2016 International Mechanical Engineering Congress \& Exposition
	(IMECE), Phoenix, AZ, November 11-–17, 2016. 

% LAMMPS package
\bibitem{Plimpton1995}
S. Plimpton, 
\href{http://www.sandia.gov/~sjplimp/papers/jcompphys95.pdf}
	{J. Comp. Phys. \textbf{117}, 1 (1995)}.

% New force and heat current formulas
\bibitem{Fan2015}
Z. Fan, L. F. C. Pereira, H. Q. Wang, J. C. Zheng, D. Donadio, and A. Harju, 
\href{http://journals.aps.org/prb/abstract/10.1103/PhysRevB.92.094301}
	{Phys. Rev. B \textbf{92}, 094301 (2015)}.

% [11]
% Uncertainty in potential parameters 
% --> uncertainty in self-diffusion coefficient and viscosity
% LJ liquid and gasous argon
\bibitem{Angelikopoulos2012}
P. Angelikopoulos, C. Papadimitriou, and P. Koumoutsakos, 
\href{http://scitation.aip.org/content/aip/journal/jcp/137/14/10.1063/1.4757266}
	{J. Chem. Phys. \textbf{137}, 144103 (2012)}.

% MD noise (different initial velocity distributions) --> uncertainty in phonon relaxation time
% --> uncertainty in thermal conductivity --> uncertainty in temperature distribution
\bibitem{Marepalli2014}
P. Marepalli, J. Y. Murthy, B. Qiu, and X. Ruan, 
\href{https://engineering.purdue.edu/NANOENERGY/publications/Marepalli_JHT_2014.pdf}
	{J. Heat Transfer \textbf{136}, 111301 (2014)}.

% uncertainty of grain boundary migration velocity from MD simulations
\bibitem{Race2015}
C. P. Race, 
\href{http://www.tandfonline.com/doi/pdf/10.1080/08927022.2014.935774}
	{Mol. Simul. \textbf{41}(13), 1069 (2015)}.

% Lennard-Jones potential parameters for argon
\bibitem{Allen1987}
M. P. Allen and D. J. Tildesley, 
\textit{Computer Simulation of Liquids} 
(Oxford University Press, New York, 1987).  

% Tersoff potential, for silicene and silicon
\bibitem{Tersoff1988}
J. Tersoff, 
\href{http://journals.aps.org/prb/abstract/10.1103/PhysRevB.37.6991}
	{Phys. Rev. B \textbf{37}, 6991 (1988)}.
	
% [16]
% Nose-Hoover thermostat
\bibitem{Nose1984}
S. Nos\'{e}, 
\href{http://scitation.aip.org/content/aip/journal/jcp/81/1/10.1063/1.447334}
	{J. Chem. Phys. \textbf{81}, 511 (1984)}.

% Nose-Hoover thermostat
\bibitem{Hoover1985}
W. G. Hoover, 
\href{http://journals.aps.org/pra/abstract/10.1103/PhysRevA.31.1695}
	{Phys. Rev. A \textbf{31}, 1695 (1985)}.

% argon melting point, experimental thermal conductivity values
\bibitem{Touloukian1970}
Y. Touloukian, P. E. Liley, and S. C. Saxena, 
\textit{Thermophysical Properties of Matter}, Vol.~\textbf{3} 
(Plenum, New York, 1970). 

% time averagin and ensemble averaging
\bibitem{Gordiz2015}
K. Gordiz, D. J. Singh, and A. Henry, 
\href{http://scitation.aip.org/content/aip/journal/jap/117/4/10.1063/1.4906957}
	{J. Appl. Phys. \textbf{117}, 045104 (2015)}.

% thermal conductivity of silicon and germanium
% T = 500 K, k_Si = 80 W/m-K, k_Ge = 33.8 W/m-K.
\bibitem{Glassbrenner1964}
C. J. Glassbrenner and G. A. Slack,
\href{http://journals.aps.org/pr/abstract/10.1103/PhysRev.134.A1058}
	{Phys. Rev. \textbf{134}, A1058 (1964)}.

% [21]
% Sampling distribution of the mean 
% Central Limit Theorem
\bibitem{Rice2007}
J. Rice, 
\textit{Mathematical Statistics and Data Analysis} (3rd ed.), 
(Duxbury Press, Belmont, CA, 2007).

% GULP package
\bibitem{Gale2003}
J. D. Gale and A. L. Rohl, 
\href{http://www.tandfonline.com/doi/abs/10.1080/0892702031000104887#.VEFnaSm2-AM}
	{Mol. Simul. \textbf{29}, 291 (2003)}.

% 
% Interfacial thermal conductance from EMD simulations
\bibitem{Chalopin2012}
Y. Chalopin, K. Esfarjani, A. Henry, S. Volz, and G. Chen, 
\href{http://journals.aps.org/prb/abstract/10.1103/PhysRevB.85.195302}
	{Phys. Rev. B \textbf{85}, 195302 (2012)}. 

\end{thebibliography}
\end{document}